\documentclass[aps,prl,reprint,superscriptaddress,longbibliography,floatfix]{revtex4-2}

\usepackage{mathrsfs}

\usepackage{soul}
\usepackage{booktabs} 
\usepackage{epsfig}
\usepackage{graphicx}
\usepackage{amsfonts}
\usepackage[figuresright]{rotating}
\usepackage{amssymb}
\usepackage{amsmath}
\usepackage{dcolumn}
\usepackage{bm}
\usepackage{xcolor}
\usepackage[colorlinks, citecolor=blue]{hyperref}
\usepackage{amsmath,amssymb,amsfonts,bm}
\usepackage{amsthm,amsmath,amssymb}
\usepackage{lipsum}
\newcommand{\RNum}[1]{\uppercase\expandafter{\romannumeral #1\relax}}
\hypersetup{linkcolor=magenta,urlcolor=blue,citecolor=blue,pdfstartview={FitH},urlcolor=blue}
\usepackage{lmodern}
\usepackage[T1]{fontenc}
\usepackage{microtype}

\usepackage{tocloft}

\usepackage{color}
\definecolor{ForestGreen}{RGB}{34, 139, 34}

\makeatletter

\newsavebox{\@brx}
\newcommand{\llangle}[1][]{\savebox{\@brx}{\(\m@th{#1\langle}\)}%
  \mathopen{\copy\@brx\kern-0.5\wd\@brx\usebox{\@brx}}}
\newcommand{\rrangle}[1][]{\savebox{\@brx}{\(\m@th{#1\rangle}\)}%
  \mathclose{\copy\@brx\kern-0.5\wd\@brx\usebox{\@brx}}}
\makeatother

\begin{document}

\title{How Long-Range Tails Reshape Non-Hermitian Spectra}

\author{Ding Gu}
\thanks{These authors contributed equally to this work.}
\affiliation{ Institute for
	Advanced Study, Tsinghua University, Beijing,  100084, China }
 
\author{Zhanpeng Fu}
\thanks{These authors contributed equally to this work.}
\affiliation{ Institute for
	Advanced Study, Tsinghua University, Beijing,  100084, China }

\author{Yu-Min Hu}
\altaffiliation{ yuminhu@pks.mpg.de }
 \affiliation{Max Planck Institute for the Physics of Complex Systems, N\"{o}thnitzer Stra{\ss}e 38, 01187 Dresden, Germany}
\affiliation{ Institute for
	Advanced Study, Tsinghua University, Beijing,  100084, China }

\author{Zhong Wang}
\altaffiliation{ wangzhongemail@tsinghua.edu.cn }
\affiliation{ Institute for
	Advanced Study, Tsinghua University, Beijing,  100084, China }

\date{\today}

\begin{abstract} 
Exponentially decaying long-range hoppings are ubiquitous in realistic tight-binding models and are often truncated to obtain a finite-range description. We show that this approximation can fail dramatically in non-Hermitian systems under open boundary conditions: an infinitesimal long-range hopping can nonperturbatively reconstruct the spectrum and eigenstates of a short-range non-Hermitian system. The mechanism is controlled by a competition between the decay length of infinitesimal long-range hoppings and the localization length of non-Hermitian skin modes, leading to a sharp transition as the decay rate is tuned. In one dimension, we show that a squeezed generalized Brillouin zone (GBZ) replaces the original GBZ of the short-ranged Hamiltonian, yielding the reconstructed open-boundary spectrum. In two or higher dimensions, we formulate a squeezed amoeba formulation describing the reconstructed spectral density. We further show that long-range hoppings can qualitatively reshape Green’s function, which can be readily detected in experiments. 
\end{abstract}

\maketitle

\emph{Introduction--}In recent years, significant progress has been made in the field of non-Hermitian physics, particularly in understanding how boundaries can greatly influence a system's properties \cite{Ashida20,Bergholtz2021RMP}. Under open boundary conditions (OBC), the non-Hermitian skin effect \cite{EdgeStates18,Chernbands18,Biorthogonal18,Anatomy19,Yokomizo2019nonbloch,Martinez18,NonXiao20,HigherKawabata20,Generalizedbulkboundary20,CorrespondenceZhang20,Okuma2020topological,Amoeba24} becomes crucial, where almost all the eigenstates become localized at the boundaries. In this sense, a non-Bloch band theory is established for finite-range non-Hermitian systems, with the standard Brillouin zone replaced by a generalized Brillouin zone (GBZ) \cite{EdgeStates18,Yokomizo2019nonbloch}. This theoretical framework plays a fundamental role in describing the novel topological and dynamical properties under OBC \cite{ChiralSong19,Probing19Longhi,Topological20Wanjura,Green21Xue,Observation21Xiao,Liouvillian21Haga,EdgeBurst22,Observation24Xiao,Observation24Zhu,Self-Healing22Longhi,xue2025nonblochedge,Real-TimeYang25}, which are drastically different from those under periodic boundary conditions (PBC).

In realistic and engineered platforms, hopping amplitudes are typically long ranged and generally decay exponentially. In Hermitian systems, replacing exponentially decaying hoppings with a finite-range tight-binding model is usually a controlled approximation, producing only quantitative corrections. However, non-Hermitian systems challenge this expectation. Because the non-Hermitian skin effect renders spectra and eigenstates highly sensitive to weak perturbations \cite{li2020critical,Okuma2019Topological,Yokomizo2021scaling,Budich2020sensor,Impurity21Li,song2024fragile,shu2024ultraspectralsensitivitynonlocal}, exponentially small hopping tails may not be innocuous. Thus, it remains unclear whether exponentially small hopping tails can significantly alter the properties of non-Hermitian systems.

In this Letter, we find that exponentially decaying long-range hoppings, even with \emph{infinitesimal} strength, can drastically alter the OBC spectrum and eigenstates of short-range non-Hermitian systems in a nonperturbative way. This nonperturbative effect arises from the competition between two length scales: one related to the exponentially decaying hoppings and the other concerning the localization of skin modes. 
We quantitatively explore this competing mechanism by developing a theoretical framework for predicting the nonperturbative spectral and eigenstate changes induced by infinitesimal long-range hoppings. Specifically,  we show how the long-range hoppings squeeze the GBZ in one dimension (1D) and alter the amoeba formulation \cite{Amoeba24} of the OBC spectral density of states (DOS) in two and higher dimensions. We further demonstrate that small long-range hoppings can significantly reshape the response (i.e., the Green's function) of non-Hermitian systems \cite{Green21Xue}, even in cases where they do not alter non-Hermitian spectra. Our theory provides a unified framework for understanding the physical properties of non-Hermitian systems with long-range hoppings.

\begin{figure}[t]
    \centering
  \includegraphics[width=8.5cm]{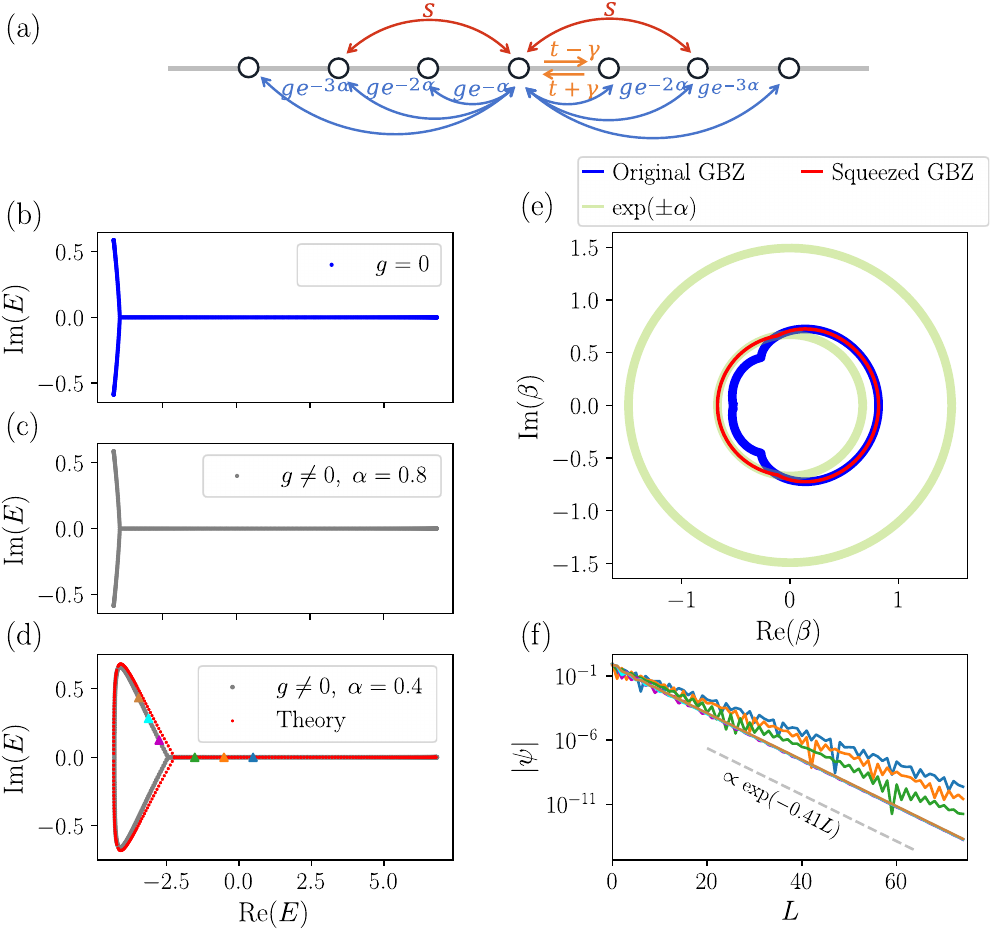}
    \caption{Squeezing of the GBZ induced by small exponentially decaying long-range hopping. 
(a) The 1D model in Eq.~\eqref{eq:1D_Ham}, with \(t=3\), \(\gamma=1\), and \(s=0.5\). (b) OBC spectrum for \(g=0\). (c) OBC spectrum for \(g=0.005\) and \(\alpha=0.8\). (d) OBC spectrum for \(g=0.005\) and \(\alpha=0.4\). Blue/gray dots in (b--d) are ED results for \(L=500\). Red dots in (d) are obtained from the squeezed GBZ via \(h_0(\beta)\), shown in (e). (e) GBZ of \(H_0\) (blue), circles \(|\beta|=e^{\pm\alpha}\) with \(\alpha=0.4\) (green), and the squeezed GBZ (red). (f) Spatial profiles of eigenstates with energies marked in (d). Eigenstates associated with the unsqueezed part of GBZ retain their original localization lengths; eigenstates associated with the squeezed part of GBZ, up to a small $O(g)$ correction, acquire an energy-independent localization length $1/\alpha$.}
    \label{fig:1D}
\end{figure}

\emph{Squeezed GBZ in  1D --}
Our main interest is to uncover the generic properties of a non-Hermitian system $H=H_0+\delta H$, where $H_0$ is a finite-range tight-binding non-Hermitian Hamiltonian and $\delta H$ describes exponentially decaying long-range hopping. Although the mechanism discussed below applies to a generic $H_0$ in arbitrary dimensions, we first illustrate it using a representative one-dimensional (1D) single-band model, as shown in Fig.~\ref{fig:1D}(a). The non-Bloch Hamiltonian of $H_0$ is
\begin{equation}
 h_0(\beta) = (t+\gamma)\beta+(t-\gamma)\beta^{-1}+s(\beta^2+\beta^{-2}).
 \label{eq:1D_Ham}
\end{equation}
Here, $\beta=e^{ik}$ defines the complexified momentum \(k\), while $h_0(e^{ik})$ with real $k\in[0,2\pi)$ gives the PBC spectrum. The exponentially decaying long-range perturbation \(\delta H\) has the non-Bloch form:
\begin{equation}
\delta h(\beta) = g\sum_{n = 1}^{L}e^{-\alpha n}(\beta^n+\beta^{-n}).
\label{eqn:1D long-range}
\end{equation}
Here, \(\alpha\) is the decay rate of the long-range hopping, \(g\) denotes its strength, and \(L\) is the system size under open boundary conditions. While our analysis applies to a small but finite \(g\), the nonperturbative spectral reconstruction is already evident for infinitesimal \(g\), provided that we take \(L\to\infty\) first and then examine \(g\to0\). This order of limits is crucial: taking \(g\to0\) at fixed \(L\) perturbatively removes long-range hoppings and smoothly recovers the finite-size spectrum of $H_0$.

Without long-range hoppings ($g=0$), the OBC spectrum of $h_0(\beta)$  in Eq.~\eqref{eq:1D_Ham} is easily obtained by non-Bloch band theory \cite{EdgeStates18,Yokomizo2019nonbloch}. With a charateristic equation $h_0(\beta)-E=0$ having $4$ roots $|\beta_1(E)|\le\cdots\le|\beta_{4}(E)|$ for a complex energy $E$, the condition $|\beta_{2}(E)|=|\beta_{3}(E)|$ between the middle two roots determines a generalized Brillouin zone (GBZ) [blue curve in Fig.~\ref{fig:1D}(e)] and the OBC spectrum [Fig. \ref{fig:1D}(b)]. The modulus of GBZ points also encodes the localization length of skin modes.

Interestingly, upon including small long-range hoppings, the OBC spectrum of $H=H_0+\delta H$ undergoes a transition when the decay rate $\alpha$ is lowered across a threshold $\alpha_c$. This transition is already visible in the limit $g\to0$. For large $\alpha$, the system remains effectively short-ranged, and its OBC spectrum is similar to the $g=0$ case [Fig.~\ref{fig:1D}(c)]. For small $\alpha$, however, slowly decaying long-range hopping substantially modifies part of the original skin modes of $H_0$, strongly reshaping the OBC spectrum [Fig.~\ref{fig:1D}(d)].

The reason why an infinitesimal $\delta H$ can induce a finite spectral change upon tuning $\alpha$ is due to the behavior of $\delta h(\beta)$ in the regimes $|\beta|<e^{-\alpha}$ or $|\beta|>e^{\alpha}$, where the summation in Eq.\eqref{eqn:1D long-range} becomes divergent when $L\to\infty$. When $\alpha$ is large, the GBZ of $H_0$ lies completely within the annulus region $e^{-\alpha}<|\beta|<e^{\alpha}$. In this case, $\delta H$ has a negligible influence on the GBZ and spectrum since for $\beta$'s on the GBZ of $H_0$, we have $\delta h(\beta)\sim g\rightarrow 0$ and $h(\beta)\approx h_0(\beta)$. This observation also indicates that the PBC spectrum of $H_0$ is always immune to an infinitesimal perturbation $\delta H$ since $e^{-\alpha}<|e^{ik}|=1<e^{\alpha}$ is automatically satisfied for a real-valued $k$. As $\alpha$ decreases below a critical value $\alpha_c\equiv\max\{|\ln|\beta||:\beta\in\text{GBZ}\}$, which is the maximum of the inverse localization lengths of all skin modes, a part of the GBZ of $H_0$ will fall outside of the annulus $e^{-\alpha}<|\beta|<e^{\alpha}$. These $\beta$'s cannot remain on the GBZ of $H$: the divergence of $\delta h(\beta)$ would lead to an unbounded energy for these skin modes, which is inconsistent with the finite spectral range observed in Fig.~\ref{fig:1D}(d). Therefore, a squeezed GBZ [red curve in Fig.~\ref{fig:1D}(e)] takes place of the original GBZ, where the parts originally outside of the annular region $e^{-\alpha}<|\beta|<e^{\alpha}$ are sequeezed to the corresponding segments on the circles $|\beta| = e^{\pm\alpha}$, while the parts originally inside the annulus remain unchanged.

\begin{figure}[t]
    \centering
    \includegraphics[width=1\linewidth]
    {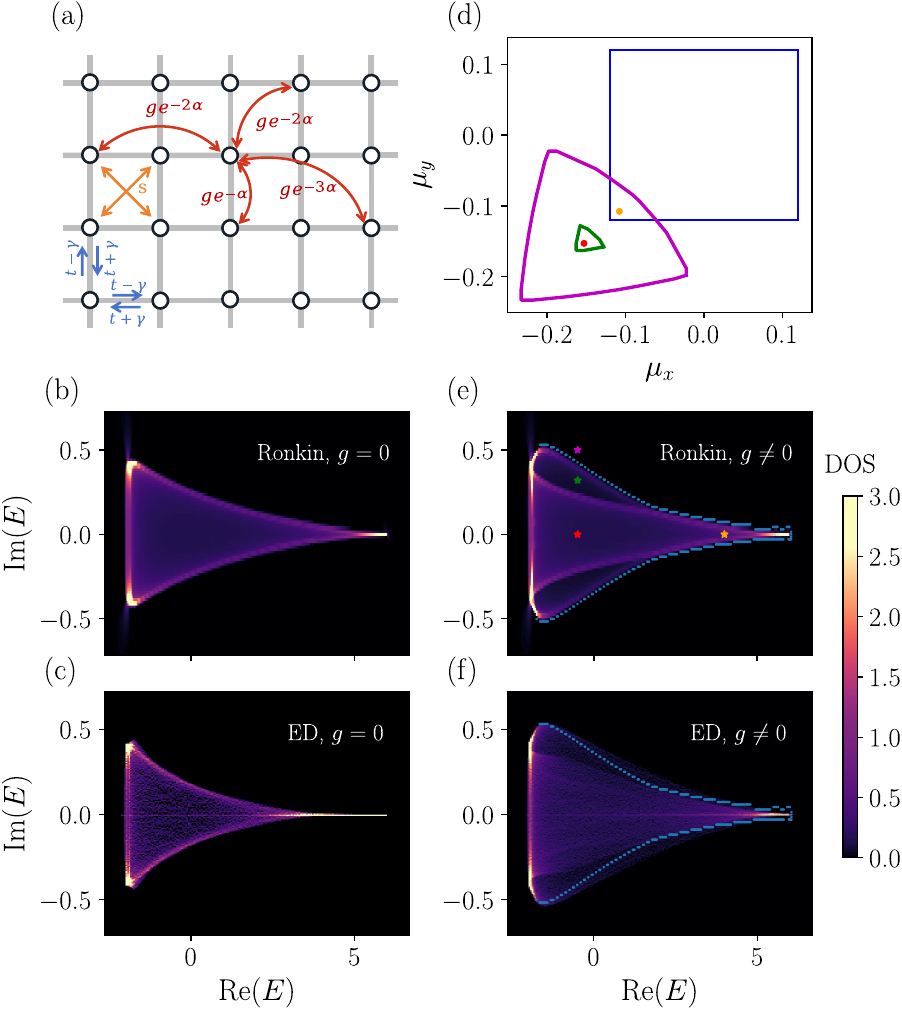}
    \caption{The density of states (DOS) of a 2D non-Hermitian system $H_0$  [Eq. \eqref{eq:2d_Ham}] subjected to small long-range hoppings $\delta H$ [Eq. \eqref{eqn:long-range hopping 2D}]. (a) The illustrative model with fixed parameters $t = 1,\gamma = 0.2,s = 0.5$. (b) DOS of $H_0$ calculated from the Ronkin function. (c) DOS of $H_0$ from exact diagonalization on a square with system size $L = 130$. An on-site random potential distributed uniformly in $[-0.5, 0.5]$ is added at each boundary site. (d)(e)(f) DOS squeezed by long-range hoppings with $\alpha = 0.12$. (d) The minimum plateaus/points of the Ronkin function $R_{E}(\mu_x,\mu_y)$ at different energies are marked by points in (e). When the Ronkin function minima are reached at a plateau, the boundaries of the plateau are plotted. (e) DOS from the squeezed amoeba formulation. (f) DOS obtained by exactly diagonalizing $H=H_0+\delta H$ on a square lattice with $g = 10^{-4}, L = 256$. We use the same boundary disorder distribution as in (c). The spectral envelope obtained from the squeezed amoeba formulation is marked in blue in (e)(f).}
    \label{fig:2D DOS}
\end{figure}

The squeezed GBZ of \(H\) can be understood as follows. To avoid a divergent energy in the thermodynamic limit, \(\beta\) must lie within the annulus \(e^{-\alpha}<|\beta|<e^{\alpha}\). In this region, \(h(\beta) = h_0+\delta h\) takes the form:
\begin{equation}\label{eq:1d_full_ham}
h(\beta) = h_0(\beta) + g\left(\frac{e^{-\alpha}\beta}{1-e^{-\alpha}\beta}
+\frac{e^{-\alpha}/\beta}{1-e^{-\alpha}/\beta}\right).
\end{equation}
Although \(\delta h(\beta)\) remains \(O(g)\) finite inside the annulus, it can still give a sizable contribution when \(\beta\) approaches the poles at \(\beta=e^{\pm\alpha}\), even for small \(g\).

Analogous to the GBZ condition \(|\beta_2(E)|=|\beta_3(E)|\) in conventional non-Bloch band theory for \(H_0\), the squeezed GBZ of \(H\), whose corresponding OBC spectrum is shown as red dots in Fig.~\ref{fig:1D}(d), is determined by pairs of roots \(\beta_a\) and \(\beta_b\) satisfying $h(\beta_a)=h(\beta_b)=E,\, |\beta_a|=|\beta_b|$. The portions of the original GBZ of \(H_0\) that lie inside the annulus remain on the GBZ of \(H\), since \(h(\beta)\approx h_0(\beta)\) there. For the newly generated segments, however, the GBZ condition cannot be satisfied if \(h(\beta)\approx h_0(\beta)\) for both roots. Therefore, one of the two roots must lie \(O(g)\) close to one of the poles \(\beta=e^{\pm\alpha}\), so that \(\delta h(\beta)\) becomes comparable to \(h_0(\beta)\). For the model in Fig. \ref{fig:1D}, we have $\beta_a=e^{-\alpha}(1+g z_a),\,
\beta_b=e^{-\alpha+i\theta}(1+g z_b)$,
with \(z_{a,b}=O(1)\). To zeroth order in \(g\), the condition \(h(\beta_a)=h(\beta_b)\) gives $
h(\beta_a)=h_0(e^{-\alpha})+\frac{1}{z_a}
= h(\beta_b)=h_0(e^{-\alpha+i\theta}).
$
Thus, in the limit \(g\to0\), the newly generated portions of the squeezed GBZ, represented by \(\beta_b\), reduce to the corresponding segments of the circles \(|\beta|=e^{\pm\alpha}\). The squeezed OBC spectrum of \(H\) is then obtained by evaluating \(h_0(\beta)\) on this GBZ, as shown by the red points in Fig.~\ref{fig:1D}(d). The squeezed GBZ also contains a small disconnected loop \cite{topologygbz25wang}, represented by \(\beta_a\), with radius \(O(g)\) near the pole \(\beta=e^{-\alpha}\). Its role is discussed further in the Supplemental Material \cite{supp}.

We further show several eigenstates in Fig. \ref{fig:1D}(f). The eigenstates associated with the unsqueezed portion of GBZ remain essentially unaffected by weak long-range hoppings. In contrast, those associated with the squeezed portion on the circle \(|\beta|=e^{-\alpha}\) are strongly modified: they collapse onto a common profile  \(|\psi(x)|\sim e^{-\alpha x}\) with an energy-independent localization length $1/\alpha$.

The squeezed GBZ theory established here generally applies to 1D non-Hermitian systems subject to weak long-range perturbations,  revealing the length-scale competition between long-range hoppings in $\delta H$ and skin modes in $H_0$. While the squeezed GBZ faithfully predicts the reconstructed spectra in the thermodynamic limit for $g\to 0$,  there are scenarios where the $O(g)$ correction (particularly the sign of $g$) is also relevant for small but finite $g$ (see the End Matter for details). 
 
\emph{DOS in 2D from squeezed-domain Ronkin function--}  It is naturally expected that the competing mechanism also emerges in higher-dimensional systems with small long-range hoppings. However, the higher-dimensional GBZ cannot be calculated directly. Instead, the OBC spectrum is encoded in an amoeba formulation of higher-dimensional non-Bloch band theory \cite{Amoeba24}. Therefore, a generalized amoeba theory compatible with long-range perturbations is needed.

To this end, we first review the standard amoeba formulation for a short-range 2D non-Hermitian Hamiltonian $H_0$ described by a non-Bloch Hamiltonian $h_0(\beta_x,\beta_y)$. For instance, the unperturbated part of the 2D model illustrated in Fig. \ref{fig:2D DOS}(a) is described by
\begin{equation}\label{eq:2d_Ham}
\begin{aligned}
h_0(\beta_x,\beta_y) & = (t+\gamma)(\beta_x+\beta_y)+(t-\gamma)(\beta_x^{-1}+\beta_y^{-1})\\
& +s(\beta_x+\beta_x^{-1})(\beta_y+\beta_y^{-1}).
\end{aligned}
\end{equation}
A central concept in the amoeba formulation is the Ronkin function, defined as a 2D integral over a torus parametrized by real variables $\theta_{x,y}\in[0,2\pi]$:
\begin{equation}\label{eq:ronkin}
R_E(\mu_x,\mu_y) = \int_{T^2}\frac{\mathrm{d}\theta_x\mathrm{d}\theta_y}{(2\pi)^2}\,\log|f(e^{\mu_x+i\theta_x},e^{\mu_y+i\theta_y})|,
\end{equation}
where $f(\beta_x,\beta_y)=\det\left[E-h_0(\beta_x,\beta_y)\right]$ is the charateristic equation for a given energy $E$. In the thermodynamic limit with generic (i.e., non-fine-tuned) open boundary conditions, the spectral density of states (DOS) $\rho(E)$ is given by $\rho(E)=\frac{1}{2\pi}\nabla^2 \phi(E)$, where a potential function $\phi(E)$ can be viewed as a 2D Coulomb potential on the complex energy plane, with the eigenvalue distribution $\rho(E)$ of $H_0$ playing the role of electric charges. In the amoeba formulation, $\phi(E)$ corresponds to the \emph{global} minimum of the Ronkin function:
\begin{align}
\phi(E) = \min_{\mu_x,\mu_y\in\mathbb{R}} R_E(\mu_x,\mu_y)
\label{eqn:DOS from Ronkin}
\end{align}

Ref.\cite{Amoeba24} has shown that $R_{E}(\mu_x,\mu_y)$ is a convex function over the entire $(\mu_x,\mu_y)$ plane, whose minimum is attained either at a single point or on a finite plateau. It has been established that if the Ronkin function minimum forms a plateau, the corresponding energy $E$ does not belong to the OBC spectrum of $H_0$. However, if the minimum is attained at a single point $(\tilde\mu_x,\tilde\mu_y)$, the spectral DOS at $E$ is nonzero and the corresponding eigenstate exhibits an asymptotic form $|\psi_E(x,y)| \sim e^{\tilde\mu_x x + \tilde\mu_y y}$. This correspondence implies that any modification of the Ronkin-function minima directly affects the spectral DOS $\rho(E)$.

Similar to the 1D case, we ask how weak long-range hoppings reshape the spectrum and eigenstates of 2D non-Hermitian systems. We consider $H=H_0+\delta H$, with non-Bloch Hamiltonian $h(\beta_x,\beta_y)=h_0(\beta_x,\beta_y)+\delta h(\beta_x,\beta_y)$, where
\begin{equation}
\delta h(\beta_x,\beta_y)
= g\sum_{n,m=-L}^{L} e^{-\alpha(|n|+|m|)}\beta_x^n\beta_y^m .
\label{eqn:long-range hopping 2D}
\end{equation}
Writing $\beta_x=e^{\mu_x+i\theta_x}$ and $\beta_y=e^{\mu_y+i\theta_y}$, we note that $\delta h(\beta_x,\beta_y)$ converges only inside the square domain $|\mu_x|<\alpha$, $|\mu_y|<\alpha$. Within this domain, $\delta h=O(g)$ and becomes negligible as $g\to0$, whereas outside it $\delta h$ diverges in the thermodynamic limit. Analogous to the 1D case, where the GBZ is confined to the annulus $e^{-\alpha}<|\beta|<e^{\alpha}$, the relevant Ronkin minimization in 2D is therefore squeezed to the square domain $|\mu_x|<\alpha$, $|\mu_y|<\alpha$, where $h(\beta_x,\beta_y)\approx h_0(\beta_x,\beta_y)$. The DOS of $H_0$ perturbed by $\delta H$ is then obtained from the squeezed Ronkin minimum
\begin{align}
\phi(E)=\min_{|\mu_x|,|\mu_y|<\alpha} R_E(\mu_x,\mu_y),
\label{eqn:DOS from squeezed Ronkin}
\end{align}
where $R_E(\mu_x,\mu_y)$ is defined as in Eq.~\eqref{eq:ronkin} using $h_0$. Under this restriction, $E$ belongs to the OBC spectrum of $H$ only when the minimum within the square domain is attained at a single point.

Figure~\ref{fig:2D DOS} presents numerical results for the Hamiltonian in Eq.~\eqref{eq:2d_Ham} with weak long-range hoppings in Eq.~\eqref{eqn:long-range hopping 2D}. When $g=0$, the DOS of $H_0$ obtained from real-space exact diagonalization [Fig.~\ref{fig:2D DOS}(c)] agrees well with that computed using the standard amoeba formulation based on Eq.~\eqref{eqn:DOS from Ronkin} [Fig.~\ref{fig:2D DOS}(b)]. Upon including the perturbation $\delta H$, the OBC spectrum is reconstructed, as shown by real-space exact diagonalization in Fig.~\ref{fig:2D DOS}(f), which is accurately captured by the squeezed amoeba formulation based on Eq.~\eqref{eqn:DOS from squeezed Ronkin} [Fig.~\ref{fig:2D DOS}(e)].

The spectral reconstruction can be understood as follows. For energies belonging to the original spectrum of $H_0$, the global Ronkin minimum in  Eq.~\eqref{eqn:DOS from Ronkin} is attained at a single point $(\tilde{\mu}_x,\tilde{\mu}_y)$, as exemplified by the yellow and red points in Fig.~\ref{fig:2D DOS}(d). If this point lies within the square domain $|\mu_x|<\alpha$ and $|\mu_y|<\alpha$ [the yellow dot in Fig.~\ref{fig:2D DOS}(d)], Eqs.~\eqref{eqn:DOS from Ronkin} and \eqref{eqn:DOS from squeezed Ronkin} yield the same local potential $\phi(E)$. Consequently, the energy $E$ remains in the spectrum of $H$, the local DOS is unchanged, and the localization length of the corresponding eigenstate is unaffected. By contrast, if $(\tilde{\mu}_x,\tilde{\mu}_y)$ lies outside this square domain [the red dot in Fig.~\ref{fig:2D DOS}(d)], the squeezed Ronkin minimum in Eq.  \eqref{eqn:DOS from squeezed Ronkin} is instead achieved at a point on the domain boundary. In this case, although $E$ continues to belong to the spectrum of $H$, the DOS at $E$ is modified due to different outcomes between Eqs.~\eqref{eqn:DOS from Ronkin} and \eqref{eqn:DOS from squeezed Ronkin}. Meanwhile, the corresponding eigenstate localization length is now determined by this new boundary minimum. For the specific model considered here, the squeezed minimum occurs at $(-\alpha,-\alpha)$, which follows from the symmetry $h(\beta_x,\beta_y)=h(\beta_y,\beta_x)$ and the convexity of $R_E(\mu_x,\mu_y)$.

For energies outside the spectrum of $H_0$, the global minima of $R_{E}(\mu_x,\mu_y)$ are attained on a plateau, as exemplified by the two extended regions enclosed by the purple and green contours in Fig.~\ref{fig:2D DOS}(d). If this plateau has a finite overlap with the square region $|\mu_x|<\alpha$ and $|\mu_y|<\alpha$ [the purple contour in Fig.~\ref{fig:2D DOS}(d)], the energy $E$ does not belong to the spectrum of $H$, since the squeezed Ronkin minimum in Eq.  \eqref{eqn:DOS from squeezed Ronkin} retains the same plateau and the corresponding potential $\phi(E)$ remains unchanged. By contrast, if the plateau has no overlap with the square region [the green contour in Fig.~\ref{fig:2D DOS}(d)], the squeezed Ronkin minimum in Eq. \eqref{eqn:DOS from squeezed Ronkin} is instead attained at a single point located on the domain boundary, which in the present model is $(-\alpha,-\alpha)$. In this case, the energy $E$ enters the squeezed spectrum of $H$, with the eigenstate localization length determined by this boundary minimum.

These results show that the squeezed amoeba formulation faithfully reproduces the OBC spectrum of a 2D non-Hermitian Hamiltonian with infinitesimal long-range hopping. In the End Matter, we consider another type of 2D long-range hopping, whose convergence domain is a disk, to demonstrate the general applicability of this formulation.

\begin{figure}[t]
    \centering
  \includegraphics[width=8.5cm]{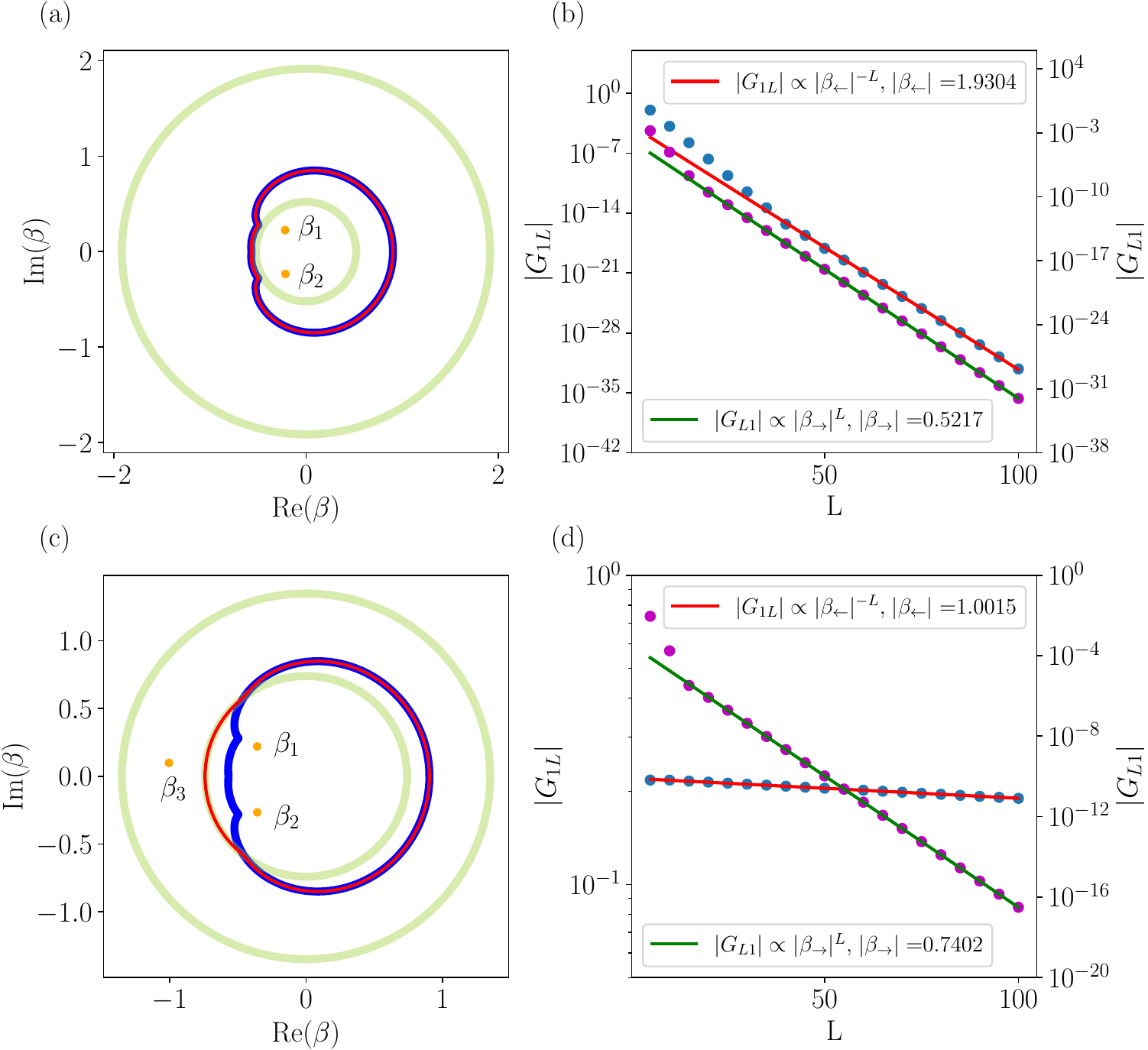}
    \caption{The scaling of the Green's function for $h_0(\beta)$ in Eq.\eqref{eq:1D_Ham} subjected to long-range hoppings $\delta h(\beta)$ in Eq.\eqref{eqn:1D long-range}. The parameters are $t = 3.0, \;\gamma = 0.5,\; s = 0.5, \;g = 0.01$. Upper panel: $\alpha=0.65,\; \omega=-6.5+0.1i$.  (a) GBZ of $h_0(\beta)$ is shown in blue, GBZ of $h(\beta)$ is depicted in red, and circles $e^{\pm \alpha}$ are shown in light green. The roots of $\omega-h_0 (\beta)= 0$ within the plot range are indicated in yellow. (b) Scaling of the Green's function: $|G_{1L}|\sim|\beta_{\leftarrow}|^{-L}$with $|\beta_{\leftarrow}| = 1.9304\approx e^{\alpha}$; $|G_{L1}|\sim|\beta_{\rightarrow}|^{L}$ with $|\beta_{\rightarrow}| = 0.5217 \approx e^{-\alpha}$.
    Lower panel: $\alpha=0.3,\; \omega=-5+0.1i$. (c) This panel is analogous to (a). (d) Scaling of the Green's function: $|G_{1L}|\sim|\beta_{\leftarrow}|^{-L}$ with $|\beta_{\leftarrow}| = 1.0015\approx |\beta_3|$; $|G_{L1}|\sim|\beta_{\rightarrow}|^{L}$with $|\beta_{\rightarrow}| = 0.7402 \approx e^{-\alpha}$.}
    \label{fig:Green}
\end{figure}

\emph{Impact on the Green's function--}We now study how infinitesimal long-range hoppings impact the experimentally relevant non-Hermitian Green's function $G(\omega)=\frac{1}{\omega-H}$ for a complex frequency $\omega$\footnote{The imaginary part of complex frequencies used in Fig. \ref{fig:Green} can be understood as an additional on-site gain/loss term in the non-Hermitian Hamiltonian $h(\beta)$.}. As an illustration, we focus on two end-to-end elements $G_{1L}(\omega)$ and $G_{L1}(\omega)$ of the OBC model described by Eq.\eqref{eq:1d_full_ham}, where $L$ is the system size of the 1D chain. 

When $g=0$,   $G_{ij}(\omega)$ of a short-range Hamiltonian $h_0(\beta)$ in Eq.~\eqref{eq:1D_Ham} can be calculated through a GBZ contour integral  \cite{Green21Xue}: 
\begin{equation}\label{eq:green_gbz}
G_{ij}(\omega) = \int_{\text{GBZ}}\frac{d\beta}{2\pi i\beta}\frac{\beta^{i-j}}{\omega-h_0(\beta)}.
\end{equation}
The residue theorem provides the asymptotic scalings $G_{L1}(\omega)\sim[\beta_2(\omega)]^L$  and $G_{1L}(\omega)\sim[\beta_{3}(\omega)]^{-L}$, where $\beta_2(\omega)$ and  $\beta_{3}(\omega)$ are the middle two of norm-ordered roots $|\beta_1(\omega)|\le\cdots\le|\beta_{4}(\omega)|$ of $\omega-h_0(\beta) = 0$.

An infinitesimal long-range perturbation $\delta h(\beta)$ gives rise to a squeezed GBZ. This leads to replacing the original GBZ contour in Eq.\eqref{eq:green_gbz} by the squeezed GBZ and changing $h_0(\beta)$ [Eq.\eqref{eq:1D_Ham}] of the denominator into $h(\beta)$ in Eq. \eqref{eq:1d_full_ham}. With $g\to0$, the roots of $\omega-h(\beta) = 0$ (i.e., the integrand's poles) consist of roots of $\omega-h_0(\beta)=0$ and two additional ones with an $O(g)$ distance to $e^{-\alpha}$ and $e^{\alpha}$. Consequently,  $G_{1L}$ and $G_{L1}$ depend on the relation between $e^{\pm \alpha}$ and the original four roots $|\beta_1|\le|\beta_2|\le|\beta_3|\le|\beta_{4}|$ of $\omega-h_0(\beta)=0$.

\begin{table}[]
\begin{tabular}{cccc}
\toprule
 Index&  Root relation&  $|G_{L1}(\omega)|$&  $|G_{1L}(\omega)|$\\ 
 \midrule 
 (1)&  $e^{-\alpha}\le|\beta_2|\le|\beta_3|\le e^{\alpha}$&  $|\beta_2|^{L}$&  $|\beta_3|^{-L}$\\ 
 (2)&  $|\beta_2|\le e^{-\alpha}\le e^{\alpha}\le|\beta_3|$&  $(e^{-\alpha})^{L}$&  $(e^{\alpha})^{-L}$\\ 
 (3)&  $|\beta_2|\le e^{-\alpha}\le|\beta_3|\le e^{\alpha}$&  $(e^{-\alpha})^{L}$&  $|\beta_3|^{-L}$\\ 
 (4)&  $e^{-\alpha}\le|\beta_2|\le e^{\alpha}\le|\beta_3|$&  $|\beta_2|^{L}$&  $(e^{\alpha})^{-L}$\\ 
 (5)&  $|\beta_2|\le|\beta_3|\le e^{-\alpha}\le e^{\alpha}$&  $(e^{-\alpha})^{L}$&  $(e^{-\alpha})^{-L}$\\ 
 (6)&  $e^{-\alpha}\le e^{\alpha}\le|\beta_2|\le|\beta_3|$&  $(e^{\alpha})^{L}$&  $(e^{\alpha})^{-L}$\\ 
 \bottomrule
\end{tabular}
\caption{The asymptotical scaling of Green's functions in non-Hermitian systems with long-range hoppings. }
\label{the_table}
\end{table}

The relative arrangement of these roots falls into six classes, whose effects on the Green's functions are summarized in Table~\ref{the_table}. In cases (1)--(4), the scaling is efficiently determined by the middle two roots among $\{\beta_1,\beta_2,\beta_3,\beta_4,e^{-\alpha},e^{\alpha}\}$, analogous to the short-range case discussed below Eq.~\eqref{eq:green_gbz}. Remarkably, cases (2)--(4) imply a possibility that long-range hoppings can substantially modify $G_{ij}(\omega)$ even for $\alpha>\alpha_c$, where the OBC spectrum and GBZ remain unchanged [see an example of case (2) in Figs.~\ref{fig:Green}(a) and \ref{fig:Green}(b)]. Furthermore, Figures~\ref{fig:Green}(c) and \ref{fig:Green}(d) illustrate case (3) with $\alpha<\alpha_c$, where the GBZ is squeezed. In this regime, for certain driving frequencies $\omega$, the Green's-function scaling is governed by $e^{-\alpha}$, the largest-modulus root inside the squeezed GBZ, and $\beta_3(\omega)$, the smallest-modulus root outside it. The last two cases, (5) and (6), imply directional amplification with a frequency-independent amplification rate $e^{\pm\alpha}$, in contrast to the frequency-dependent amplification in Ref.~\cite{Green21Xue}. The detailed analysis of these two cases involves additional technical considerations and is therefore deferred to the Supplemental Material~\cite{supp}.
In summary, the long-range hoppings set $\alpha$ as a bound for the possible spatial decay rates of Green's functions.

{\it Discussion--} We have established a generic framework for understanding how weak exponentially decaying long-range hoppings can affect the eigenstates, spectra, and dynamic responses of non-Hermitian systems in a nonperturbative way. The squeezed amoeba formulation, reducing to the squeezed GBZ theory in 1D non-Hermitian systems \cite{supp}, has broad applicability in predicting the physical properties of long-range non-Hermitian systems. While the illustrative examples are non-Hermitian Hamiltonians, our theory also applies to Lindbladian open quantum systems (see Supplemental Material \cite{supp}). 

{\it Acknowledgments--}This work is supported by the National Natural Science Foundation of China (Grants No. 12125405), the National Key R\&D Program of China (No. 2023YFA1406702), and Quantum Science and Technology-National Science and Technology Major Project (Grant No. 2021ZD0302502).

\bibliography{references}
\newpage
\appendix
\section*{End Matter}

\begin{figure}[h]
    \centering
    \includegraphics[width=8.5cm]{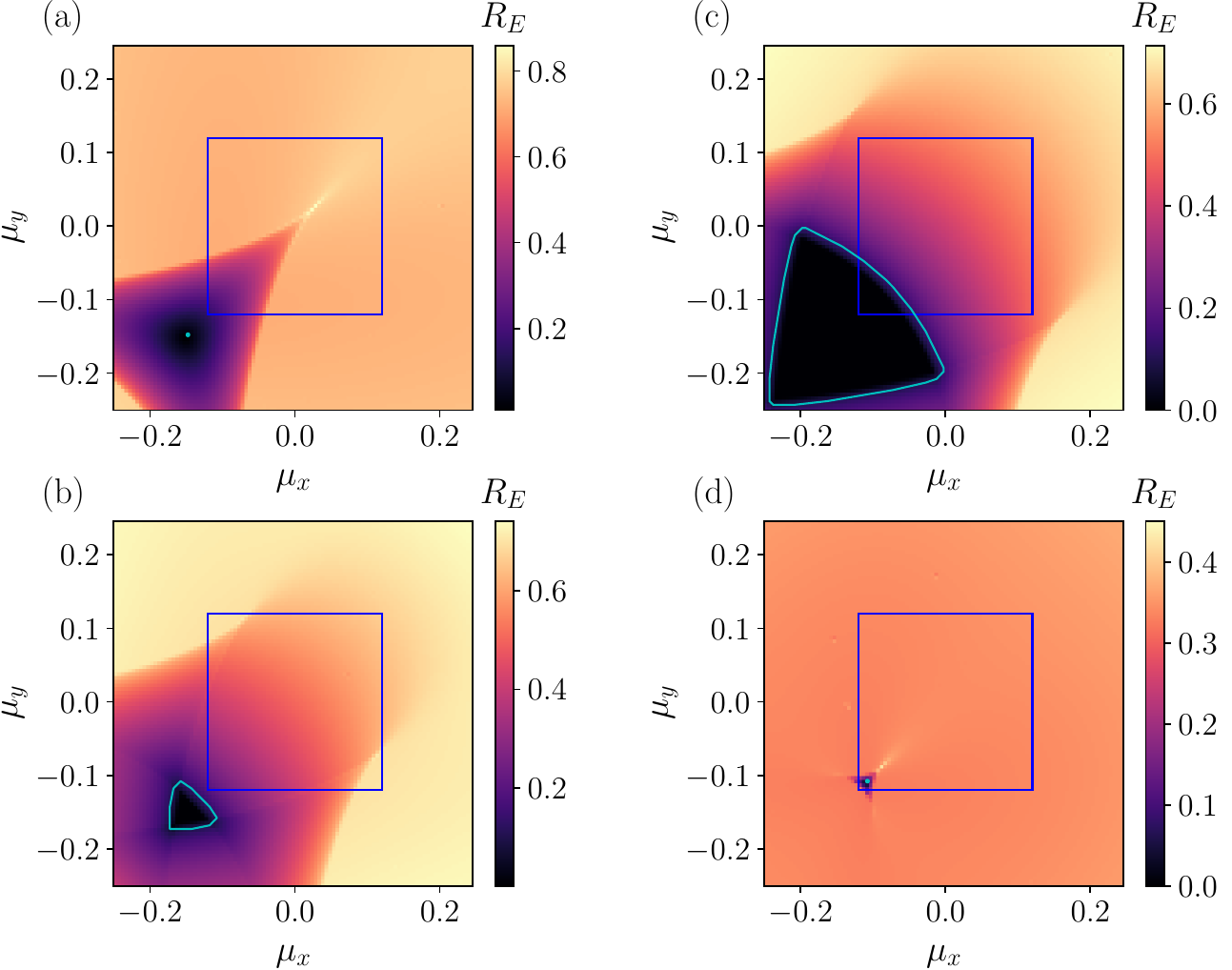}
    \caption{Ronkin function $R_{E}(\mu_x,\mu_y)$ of different energies on the $(\mu_x, \mu_y)$ plane. The blue square corresponds to the boundary of the restricted domain $|\mu_x|, |\mu_y| \leq\alpha$. (a)(d) The minimum of the Ronkin function is reached at a single point. (b)(c) The minimum of the Ronkin function is reached on a plateau.      }
    \label{fig:ronkin_func}
\end{figure}

\emph{Details of the Ronkin function.--} In Fig.~\ref{fig:ronkin_func}, we show the values of the Ronkin function $R_{E}(\mu_x,\mu_y)$ on the $(\mu_x,\mu_y)$ plane. The energies $E$'s are taken as those illustrated in Fig.~\ref{fig:2D DOS}(d)(e). Both the restricted domain and the minimum of the Ronkin function are marked. As discussed in the main context, an energy $E$ is in the spectrum only if the minimum of the corresponding Ronkin function within the restricted domain is reached at a single point. Consequently, only the energy corresponding to Fig.~\ref{fig:ronkin_func} (c) does not belong to the squeezed spectrum.

\begin{figure}[t]
    \centering
    \includegraphics[width=8.5cm]{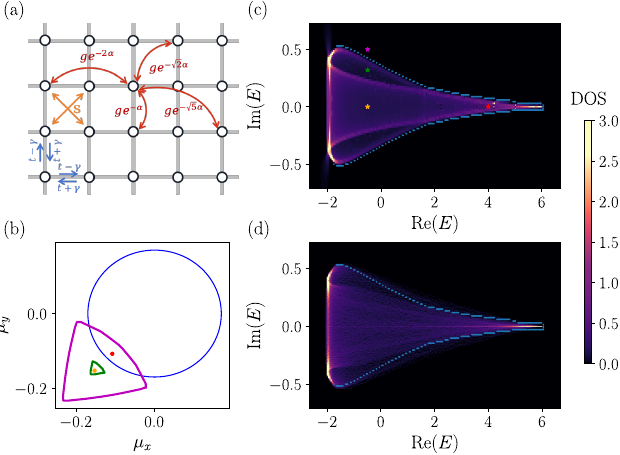}
    \caption{DOS in a 2D system with 2-norm exponentially decaying hopping. (a) The model with long-range hopping decaying as $g\exp(-\alpha\sqrt{(r_x-r'_x)^2+(r_y-r_y')^2})$. (b) Relative positions between the Ronkin function minima and the restricted Ronkin function domain. The blue circle marks the boundary of the restricted region $\mu_x^2+\mu_y^2<\alpha^2$. For energies marked with different colors in (c), if the Ronkin function minimum is reached at a single point, the point is plotted with the same color. If the Ronkin function minimum is reached at a plateau, the boundary of the plateau is plotted with the same color. (c) and (d) show DOS derived from the Ronkin function and ED, respectively. The DOS envelope is highlighted in blue, where the results from the Ronkin function agree with those from ED. The parameters are $t=1, \gamma=0.2, s=0.5, \alpha=0.17, g=10^{-4}, L = 256$. In ED, we add an on-site random potential distributed uniformly in $[-0.5, 0.5]$ at each boundary site of the square lattice.} 
    \label{fig:2-norm}
\end{figure}

\emph{A 2D system with 2-norm exponentially decaying hoppings.--}In the main text, we focused on the case where the strength of long-range hopping decays exponentially in the 1-norm distance Eq.~(\ref{eqn:long-range hopping 2D}). Here, we consider a more common form of exponentially decaying hopping:
$\delta h(\beta_x, \beta_y) = g\sum_{n,m=-L}^{L}e^{-\alpha\sqrt{n^2+m^2}}\beta_x^n\beta_y^m$
, where the hopping strength depends on the 2-norm distance. The unperturbed Hamiltonian $h_0(\beta_x,\beta_y)$ follows the same form as Eq.\eqref{eq:2d_Ham} in the main context. We denote $\beta_{x,y}$ as $e^{\mu_{x,y}+i\theta_{x,y}}$. $\delta h(\beta_x, \beta_y)$ now diverges when $\mu_x^2+\mu_y^2>\alpha^2$, while it becomes negligible $\delta h(\beta_x, \beta_y)\sim 0$ within the circular region $\mu_x^2+\mu_y^2<\alpha^2$. Consequently, the domain of the Ronkin function is restricted to the circular region $\mu_x^2+\mu_y^2<\alpha^2$, with the boundary highlighted in blue in Fig.~\ref{fig:2-norm}(b). The potential function $\phi(E)$ that determines the spectral DOS $\rho(E)$ is defined as $\phi(E) = \min_{\mu_x^2+\mu_y^2<\alpha^2} R_E(\mu_x,\mu_y)$.

Fig.~\ref{fig:2-norm}(c) and (d) show the density of states derived from the squeezed amoeba formulation and from exact diagonalization, respectively, which clearly indicates the agreement between the two results. In Fig.~\ref{fig:2-norm}(b), the Ronkin minima for different energies, indicated by various colors in Fig.~\ref{fig:2-norm}(c), are displayed.

\emph{Finite $g$ correction.--} We have discussed in the main text that exponentially decaying long-range hoppings with infinitesimal strength $g$ can induce $O(1)$ changes in the system's spectrum, eigenstates, and GBZ. The theoretical predictions of these $O(1)$ renormalizations agree well with numerical results in both 1D and 2D systems. However, in certain scenarios, the $O(g)$ contributions from long-range hoppings are also significant. In this section, we discuss in detail how these $O(g)$ corrections manifest, using specific examples in 1D and 2D.

\begin{figure}[t]
    \centering
    \includegraphics[width=0.75\linewidth]{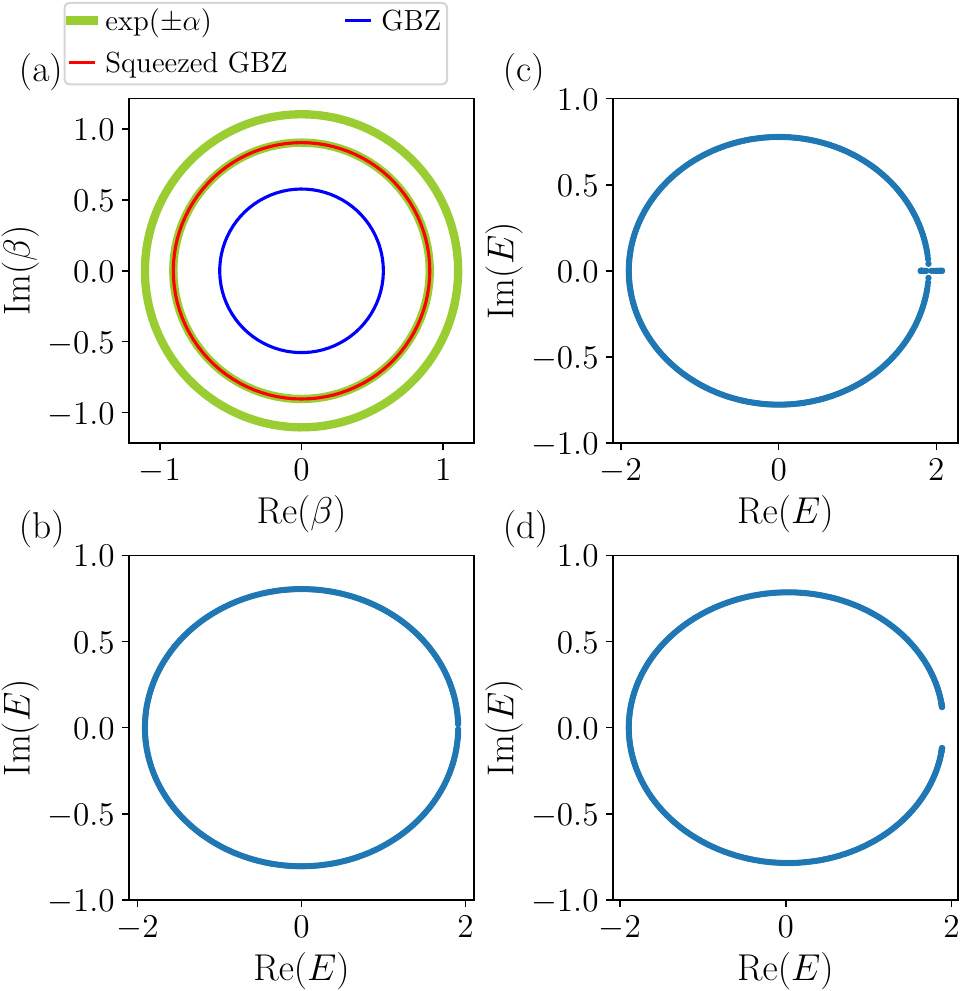}
    \caption{GBZ and spectrum for the Hatano-Nelson model with long-range perturbations. (a) Green circles are $|\beta| = e^{\pm\alpha}$. The blue circle is GBZ of the $h_0$: $|\beta|=\sqrt{(t-\gamma)/(t+\gamma)}$. The red circle is the squeezed GBZ: $|\beta| = e^{-\alpha}$. The parameters are $t=1.0$, $\gamma=0.5$, $\alpha = 0.1$. (b) Spectrum obtained from the squeezed GBZ. (c)(d) Spectra from ED at $L = 500$ for long-range hopping strength $g = 0.005$ and $g = -0.005$, respectively. The nuremcial results differ from the squeezed GBZ prediction near $\beta = e^{-\alpha+i\theta},\,\theta\approx 0$.}
    \label{fig:spec_1d app}
\end{figure}

In 1D, we illustrate this with a simple model $h(\beta) = h_0(\beta)+\delta h(\beta)$,
where $h_0(\beta)  = (t+\gamma)\beta+(t-\gamma)\beta^{-1}$, whose GBZ is a circle $|\beta| = \sqrt{(t-\gamma)/(t+\gamma)}$. Similarly, $\delta h(\beta)  = g\sum_{n = 1}^Le^{-n\alpha}(\beta^n+\beta^{-n})$. When $\delta h$ is added, the renormalization of the GBZ is straightforward: the GBZ is unchanged when $e^{-\alpha} < \sqrt{(t-\gamma)/(t+\gamma)}$ and is replaced by the circle $|\beta| = e^{-\alpha}$ when $e^{-\alpha} > \sqrt{(t-\gamma)/(t+\gamma)}$, as shown in Fig.~\ref{fig:spec_1d app} (a). The spectrum computed from the squeezed GBZ through $h_0(\beta)$ is shown in Fig.~\ref{fig:spec_1d app} (b).

\begin{figure}[t]
    \centering
    \includegraphics[width=1\linewidth]{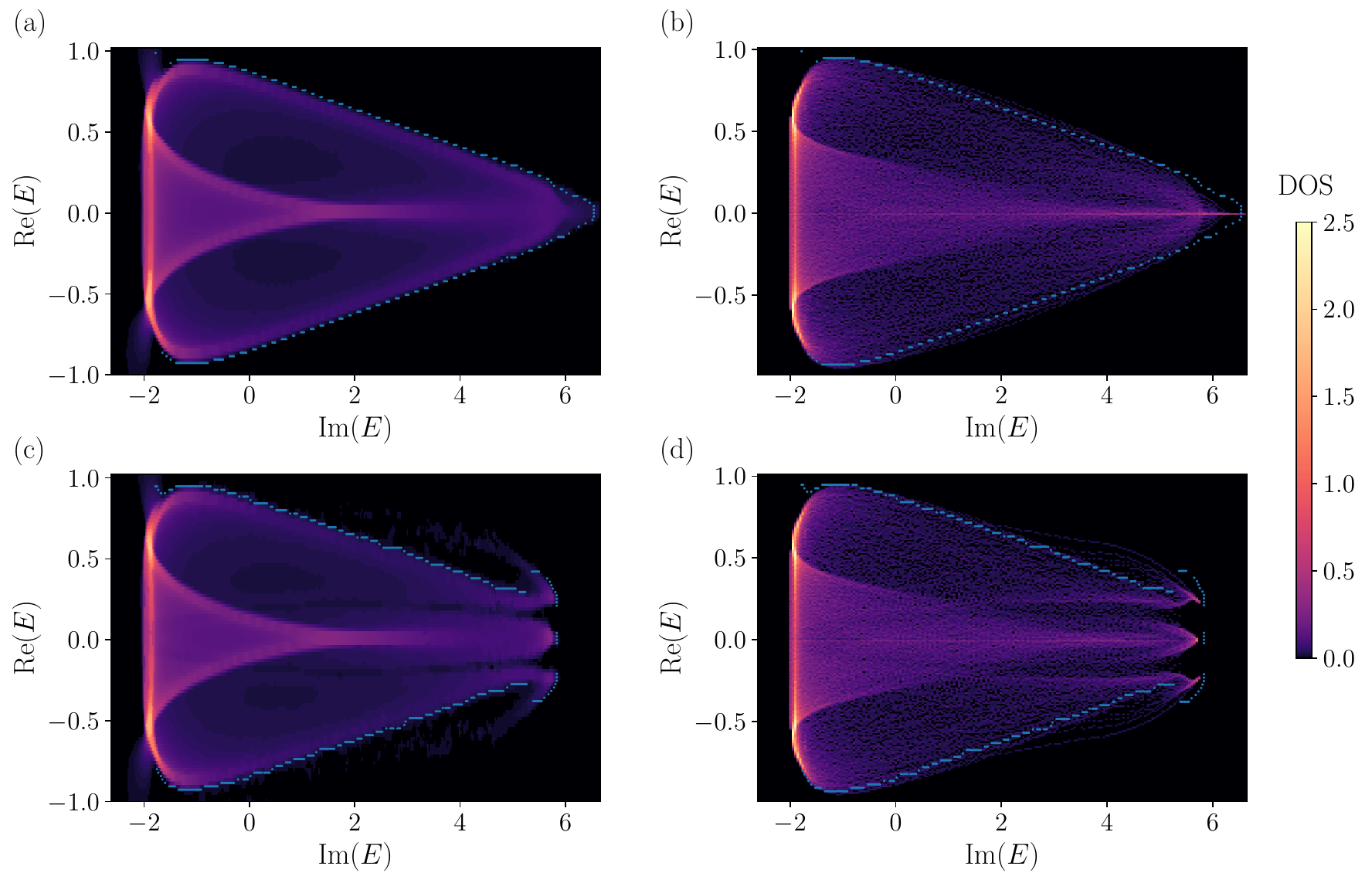}
    \caption{DOS in a 2D system where $O(g)$ correction cannot be neglected. The strength of long-range hopping decays as $g \exp(-\alpha(|r_x-r'_x|+|r_y-r_y'|))$. (a), (b) show the DOS with $g = 0.005$ derived from the Ronkin function and ED, respectively. (c), (d) show the DOS with $g = -0.005$ derived from the Ronkin function and ED respectively. The blue curve highlights the DOS envelope, which shows a great agreement between the Ronkin and ED results. $h_0(\beta_x,\beta_y)$ is the same as Eq.\eqref{eq:2d_Ham} the main context. The parameters are $t = 1.0, \gamma=0.3, s=0.5, \alpha=0.1, |g|=0.005$. The ED results are obtained with an on-site random potential distributed uniformly in $[-0.5, 0.5]$ at each boundary site. The system size in ED is $L=256$.
    }
    \label{fig:DOS_EM}
\end{figure}
The spectra derived from exact diagonalization with small but finite $g$ are illustrated in Fig.~\ref{fig:spec_1d app} (c)(d), where $g>0$ in (c) and $g<0$ in (d). The numerical results agree well with the prediction from squeezed GBZ except at the rightmost part of the spectrum, which corresponds to points on the squeezed GBZ $\beta = e^{-\alpha+i\theta}$ with $\theta\approx 0$. This discrepancy is explained as follows.

In the determination of eigenstates for 1D non-Hermitian systems, we need two $\beta$'s that satisfy: $|\beta_a| = |\beta_b|,\,h(\beta_a) = h(\beta_b)$, where $h(\beta)$ can be expressed as Eq.~(\ref{eq:1d_full_ham}). $\delta h(\beta)$ diverges when $|\beta|<e^{-\alpha}$ or $|\beta|>e^{\alpha}$ and $\delta h\sim g\rightarrow 0$ inside the annulus $e^{-\alpha}<|\beta|<e^{\alpha}$.

When \(e^{-\alpha}>\sqrt{(t-\gamma)/(t+\gamma)}\), the GBZ of \(h_0(\beta)\) lies outside the annulus \(e^{-\alpha}<|\beta|<e^{\alpha}\). Consequently, to find two roots of equal modulus inside the annulus with the same energy under \(h(\beta)\), the two roots cannot both satisfy \(h(\beta)\approx h_0(\beta)\). Instead, one of the roots, denoted by \(\beta_a\), must lie very close to the pole at \(\beta=e^{-\alpha}\), where \(\delta h(\beta)\) can contribute at \(O(1)\) to \(h(\beta)\). We may therefore write \(\beta_a=e^{-\alpha}(1+g z_a)\) and \(\beta_b=e^{-\alpha+i\theta}(1+g z_b)\), with \(z_a,z_b=O(1)\). To leading order in \(g\), these two roots satisfy \(h(\beta_a)=h_0(e^{-\alpha})+1/z_a\) and \(h(\beta_b)=h_0(e^{-\alpha+i\theta})\).We retain only $O(1)$ terms in $h(\beta_{a,b})$. $h(\beta_a) = h(\beta_b)$ leads to $1/z_a = h_0(e^{-\alpha+i\theta})-h_0(e^{-\alpha})$. When $\theta$ goes near 0, $z_a$ grows large and the $O(g)$ terms in $h(\beta)$ becomes non-negligible. This explains the deviation of ED results from the spectrum calculated from the squeezed GBZ, where we consider only the $O(1)$ approximation. Different results for different signs of $g$ are straightforward from the above analysis. For the model discussed in the main context, the $O(g)$ correction is irrelevant since no points on the squeezed GBZ are close to $\beta = e^{\pm\alpha}$.

Similar phenomena also occur in 2D. To the lowest order, the spectra for 2D systems are obtained by restricting the Ronkin function to a domain where $\delta h(\beta_x,\beta_y)$ does not diverge, and we approximate $h(\beta_x,\beta_y)\approx h_0(\beta_x,\beta_y)$. Using Eq.\eqref{eq:2d_Ham}(\ref{eqn:long-range hopping 2D}) as an example, the restricted domain is given by $|\mu_{x/y}|<\alpha$. Inside this square region:
\begin{align}
&\delta h = g(\frac{1}{1-e^{-\alpha}\beta_x}+\frac{1}{1-e^{-\alpha}\beta_x^{-1}})(\beta_x\leftrightarrow \beta_y).
\label{eqn:long-range hopping 2D app}
\end{align}
However, when the Ronkin function minimum is reached near $(\mu_x,\mu_y) = (\pm\alpha,\pm\alpha)$, $\delta h$ can not be neglected, since the denominator in Eq.\eqref{eqn:long-range hopping 2D app} approaches 0 near these points. By including  $\delta h(\beta_x,\beta_y)$ in the calculation of the Ronkin function $R_E(\mu_x,\mu_y)$ (replacing $h_0$ by $h=h_0+\delta h$), we can obtain $O(g)$ corrections that agree well with numerics.

In Fig.~\ref{fig:DOS_EM}(a)(b), we show DOS for 2D systems with $g>0$. And in Fig.~\ref{fig:DOS_EM}(c)(d), DOS for 2D systems with $g<0$ are plotted. The left and right panels illustrate results from the Ronkin function $R_{E}(\mu_x,\mu_y)$(for $h_0+\delta h$) and exact diagonalization, respectively. The major difference between different signs of $g$ lies at the rightmost part of the DOS, where it is easy to verify that the energies correspond to $(\beta_x,\beta_y)\approx (e^{-\alpha},e^{-\alpha})$.

\end{document}


\title{Supplemental Material: How Long-Range Tails Reshape Non-Hermitian Spectra}

\author{Ding Gu}
\thanks{These authors contributed equally to this work.}
\affiliation{ Institute for
	Advanced Study, Tsinghua University, Beijing,  100084, China }
 
\author{Zhanpeng Fu}
\thanks{These authors contributed equally to this work.}
\affiliation{ Institute for
	Advanced Study, Tsinghua University, Beijing,  100084, China }

\author{Yu-Min Hu}
\altaffiliation{ yuminhu@pks.mpg.de }
 \affiliation{Max Planck Institute for the Physics of Complex Systems, N\"{o}thnitzer Stra{\ss}e 38, 01187 Dresden, Germany}
\affiliation{ Institute for
	Advanced Study, Tsinghua University, Beijing,  100084, China }

\author{Zhong Wang}
\altaffiliation{ wangzhongemail@tsinghua.edu.cn }
\affiliation{ Institute for
	Advanced Study, Tsinghua University, Beijing,  100084, China }

\date{\today}

\maketitle

\section{\label{sec: green_func_sm} More on the Green's function}
In this section, we discuss further the Green's functions in non-Hermitian systems with exponentially decaying long-range hoppings.

For a finite-range non-Hermitian Hamiltonian \(H\), the Green's function \(G_{ij}(\omega)\) can be expressed as a contour integral over the GBZ \cite{Green21Xue}:
\begin{equation}\label{seq:green_gbz}
G_{ij}(\omega) = \int_{\text{GBZ}}\frac{d\beta}{2\pi i\beta}\frac{\beta^{i-j}}{\omega-h(\beta)},
\end{equation}
where \(h(e^{ik})\) denotes the Bloch Hamiltonian of \(H\) under periodic boundary conditions.

By the residue theorem, the asymptotic behavior is \(G_{L1}(\omega)\sim[\beta_M(\omega)]^L\) and \(G_{1L}(\omega)\sim[\beta_{M+1}(\omega)]^{-L}\), where \(\beta_M(\omega)\) and \(\beta_{M+1}(\omega)\) are the middle two roots among the \(2M\) roots of \(\omega-h(\beta)=0\), ordered by magnitude as
\[
|\beta_1(\omega)|\le \cdots \le |\beta_{2M}(\omega)|.
\]
Here, \(M\) roots lie inside the GBZ, and the other \(M\) roots lie outside. Accordingly, \(G_{L1}(\omega)\) is controlled by the root of largest modulus inside the GBZ, whereas \(G_{1L}(\omega)\) is controlled by the root of smallest modulus outside the GBZ.

\begin{figure}[t]
    \centering
    \includegraphics[width=0.9\linewidth]{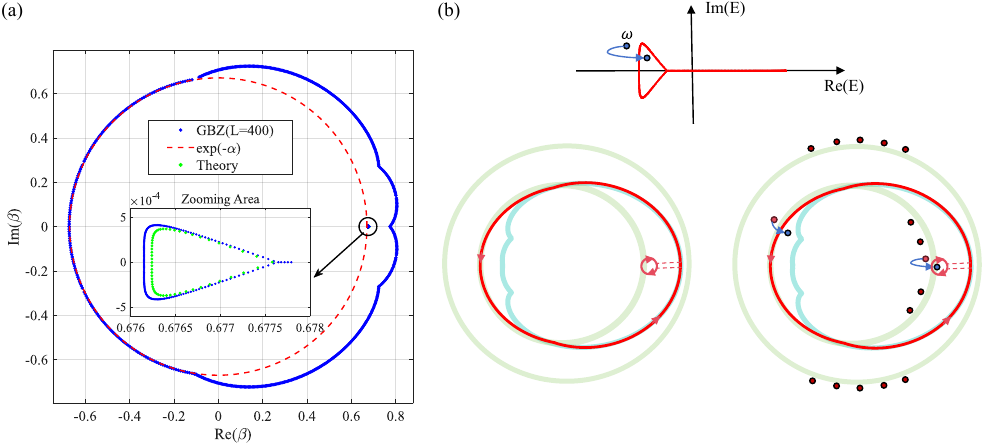}
    \caption{squeezed GBZ for finite \(g\) and schematic illustration of the change in the roots relative to the squeezed GBZ. 
(a) Numerical results for the GBZ of the model defined in Eq.~ (3) of the main text. The long-range hopping squeezes the GBZ of the system, and a small loop appears near \(\beta=e^{-\alpha}\). The size of this loop (shown in the zoomed-in region) is proportional to the long-range hopping strength \(g\) in the limit \(g\to 0\). The theoretical result is given by Eq.~\eqref{eq:theo_small_loop}. The parameters are $t = 3.0, \gamma = 1.0, s=0.5, g=-0.1, \alpha = 0.4$. The system size is $L = 400$. 
(b) The orientations of the two components of the squeezed GBZ determine its interior and exterior. When the frequency of the external drive, \(\omega\), enters the circular region enclosed by the spectrum, \(\beta_3\)---one of the roots of \(\omega-h(\beta)=0\) (red dots) lying within the annular region---crosses the GBZ from the exterior to the interior. At the same time, another root near \(e^{-\alpha}\) crosses the small loop from the interior to the exterior of the GBZ, ensuring that the winding of \(h(\beta)\) traced over GBZ around $\omega$ remains unchanged. As a result, the root with the smallest modulus outside the GBZ and the root with the largest modulus inside the GBZ have approximately the same modulus, \(|\beta|=e^{-\alpha}+\mathcal{O}(g)\). This leads to the asymptotic scalings \(G_{L1}\sim e^{-\alpha L}\) and \(G_{1L}\sim (e^{-\alpha})^{-L}=e^{\alpha L}\).}
    \label{fig:gbz_combine}
\end{figure}

To apply this formula to the long-range model, it is convenient to first consider a finite-range version of the infinite-range model at finite \(g\), and then take the limit \(g\to 0\), \(M\to\infty\):
\begin{align}
&h(\beta) = h_0(\beta)+\delta h(\beta),\nonumber\\
&\delta h(\beta) = g\sum_{n = 1}^M e^{-\alpha n}(\beta^n+\beta^{-n}).
\end{align}
Here, \(M\) denotes the range of the exponentially decaying hopping. In the limit \(g\to 0\), \(M\to\infty\), the GBZ is squeezed in the manner described in the main text: starting from the GBZ of \(h_0\), the portions lying outside the annulus \(e^{-\alpha}<|\beta|<e^{\alpha}\) are replaced by the corresponding segments of the circles \(|\beta|=e^{\pm \alpha}\), while the portions inside the annulus remain unchanged. However, for finite \(g\) and sufficiently large \(M\), the squeezed GBZ contains an additional component. For \(h_0(\beta)\) of the form discussed in the main text,
\begin{equation}
 h_0(\beta) = (t+\gamma)\beta+(t-\gamma)\beta^{-1}+s(\beta^2+\beta^{-2}),
\end{equation}
this additional component is a small loop of radius \(O(g)\) near the point \(e^{-\alpha}\). Points on this loop are paired with points on the segment of the circle \(|\beta| = e^{-\alpha}\) to which the GBZ of \(h_0\) is squeezed:
\begin{align}
\label{eq:theo_small_loop}
&\beta_a = e^{-\alpha}(1 + g z_a),\quad \beta_b = e^{-\alpha + i\theta}(1 + g z_b),\nonumber\\
&h(\beta_a) = h_0(e^{-\alpha}) + \frac{1}{z_a}
= h(\beta_b) = h_0(e^{-\alpha + i\theta}).
\end{align}
A detailed discussion of this pairing is given in the main text. In the limit \(g\to 0\), the radius of the loop shrinks to zero. However, for the analysis of the Green's function, it is necessary to keep \(g\) small but finite. Consequently, this small loop near \(e^{-\alpha}\) must be included in the integration contour when evaluating the Green's function. In Fig.~\ref{fig:gbz_combine}(a), we numerically demonstrate the existence of the small loop near \(\beta=e^{-\alpha}\) in the squeezed GBZ for finite but small \(g\). In the lower-left part of Fig.~\ref{fig:gbz_combine}(b), we schematically illustrate the integration contour, which includes both the large loop and the small loop of the squeezed GBZ.

Now that the GBZ has been obtained for finite \(g\) and sufficiently large \(M\), the remaining task is to determine the roots of \(\omega-h(\beta)=0\). The equation \(\omega-h_0(\beta)=0\) has four roots, ordered by magnitude as
\[
|\beta_1|<|\beta_2|<|\beta_3|<|\beta_4|.
\]
Because \(|\delta h(\beta)|\ll |h_0(\beta)|\) inside the annulus \(e^{-\alpha}<|\beta|<e^{\alpha}\), while \(|\delta h(\beta)|\gg |h_0(\beta)|\) outside this region, the roots of \(\omega-h_0(\beta)=0\) that lie inside the annulus remain, to a good approximation, roots of \(\omega-h(\beta)=0\). The remaining roots of \(\omega-h(\beta)=0\) must then have moduli very close to \(e^{-\alpha}\) or \(e^{\alpha}\). Altogether, the equation \(\omega-h(\beta)=0\) has \(2M\) roots. The asymptotic scaling of the Green's functions \(G_{1L}(\omega)\) and \(G_{L1}(\omega)\) then depends on the relative positions of these roots with respect to the squeezed GBZ.

The relative positions of \(\beta_{1,2,3,4}\) with respect to the squeezed GBZ are unambiguous. The subtlety arises for the roots whose moduli are close to \(e^{\pm\alpha}\). For most values of \(\omega\), the roots with moduli close to \(e^{-\alpha}\) lie inside the squeezed GBZ of \(h(\beta)\), whereas those with moduli close to \(e^{\alpha}\) lie outside it. The scenarios discussed in the main text belong to this case. Accordingly, the scaling of \(G_{L1}(\omega)\) is determined by the largest among \(|\beta_1|, |\beta_2|,\) and \(e^{-\alpha}\), while the scaling of \(G_{1L}(\omega)\) is determined by the smallest among \(|\beta_3|, |\beta_4|,\) and \(e^{\alpha}\). 

The situation changes when \(\omega\) crosses the spectrum and enters the circular region enclosed by it, as shown in Fig.~\ref{fig:gbz_combine}(b). In terms of the roots, this corresponds to \(\beta_3\) crossing the squeezed GBZ from outside to inside. Since the total number of roots inside the squeezed GBZ must remain unchanged \footnote{The spectrum traced out by \(h(\beta)\) along the squeezed GBZ should have trivial winding around any \(\omega\), and this winding is determined by the number of roots enclosed by the squeezed GBZ.}, one of the roots with modulus approximately equal to \(e^{-\alpha}\) must cross the squeezed GBZ from inside to outside. This crossing occurs on the smaller loop: when \(\beta_3\) crosses, the part of spectrum traced out by \(h(\beta)\) along the larger loop of the squeezed GBZ acquires winding number \(+1\) around \(\omega\); the subsequent crossing on the smaller loop contributes winding number \(-1\) associated with spectrum winding along the small loop. As a result, the total winding of the spectrum along the full squeezed GBZ remains trivial both before and after the crossing.

\begin{figure}[t]
    \centering
    \includegraphics[width=0.6\linewidth]{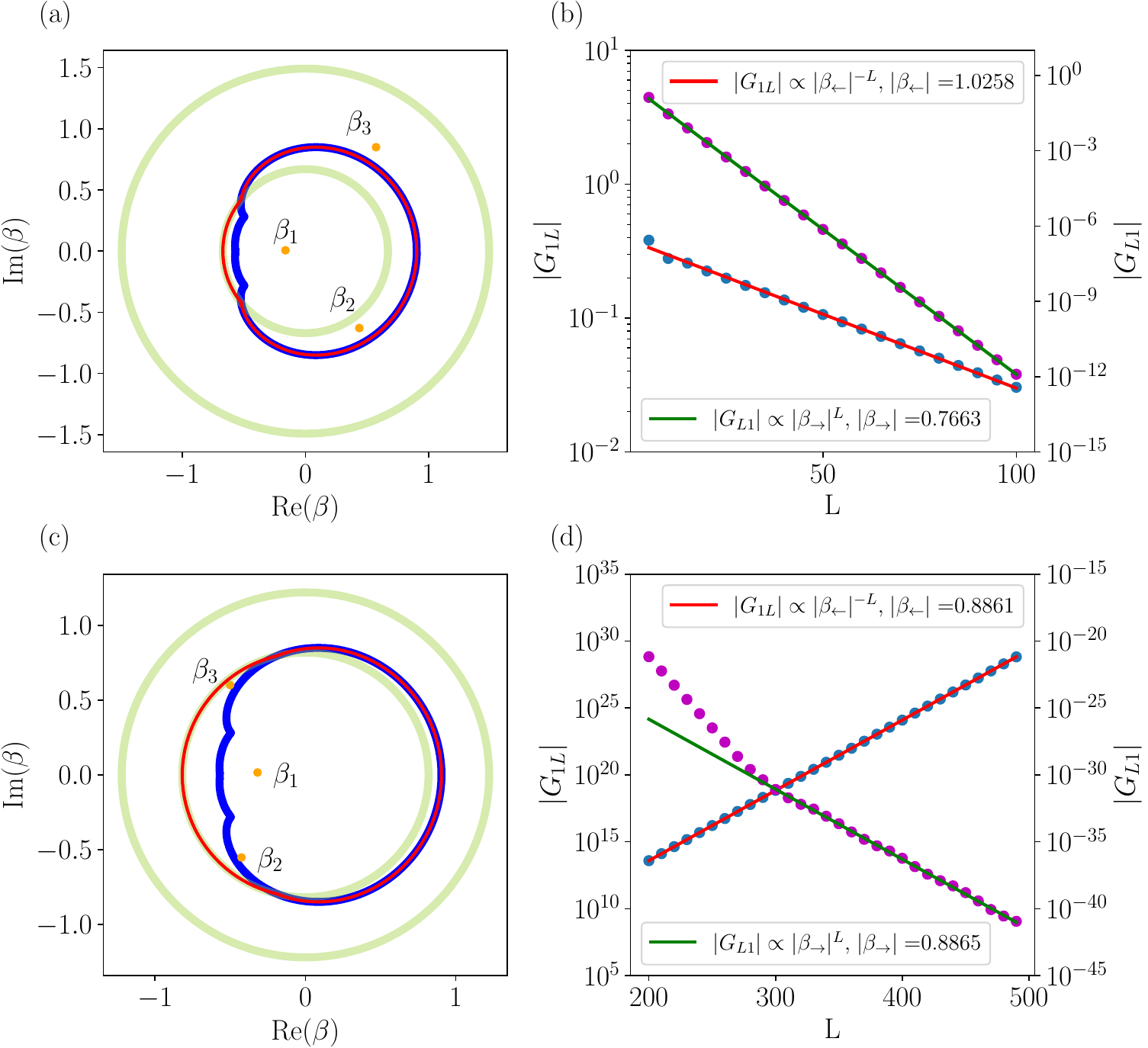}
    \caption{Scaling of the Green function of two circumstances. The parameters are $t = 3.0, \;s = 0.5,\; \gamma = 0.5, \;g = 0.01$. Upper panel: $\alpha=0.4,\; \omega=3+i$. (a), (b) show the case when the roots $\beta_2$ and $\beta_3$ of $\omega-h_0(\beta)$ lie within the ring $r\in[e^{-\alpha}, e^{\alpha}]$. The two roots of $\omega-h(\beta)$ closest to GBZ will remain approximately equal to  $\beta_2$ and $\beta_3$. As shown in (b), $|\beta_{\leftarrow}|=1.0258\approx|\beta_3|, |\beta_{\rightarrow}| = 0.7663\approx|\beta_2|$. Lower panel: $\alpha=0.12,\; \omega=-4+0.15i$. (c)
    (d) show the case where $\beta_3$ crosses the squeezed GBZ from outside to inside. As shown in (d), $|\beta_{\leftarrow}|=0.8861\approx e^{-\alpha}, |\beta_{\rightarrow}| = 0.8865\approx e^{-\alpha}$.}
    \label{fig:GF_EM}
\end{figure}

Consequently, for these values of \(\omega\), both \(G_{L1}(\omega)\) and \(G_{1L}(\omega)\) are determined by roots with modulus \(e^{-\alpha}\):
\[
|G_{L1}(\omega)|\sim e^{-\alpha L},\qquad
|G_{1L}(\omega)|\sim (e^{-\alpha})^{-L}=e^{\alpha L},
\]
thereby exhibiting directional amplification \cite{Green21Xue}. We explicitly demonstrate this scaling in Fig.~\ref{fig:GF_EM}(c,d), where the position of \(\beta_3\) is also marked.

\begin{figure}
    \centering
    \includegraphics[width=0.6\linewidth]{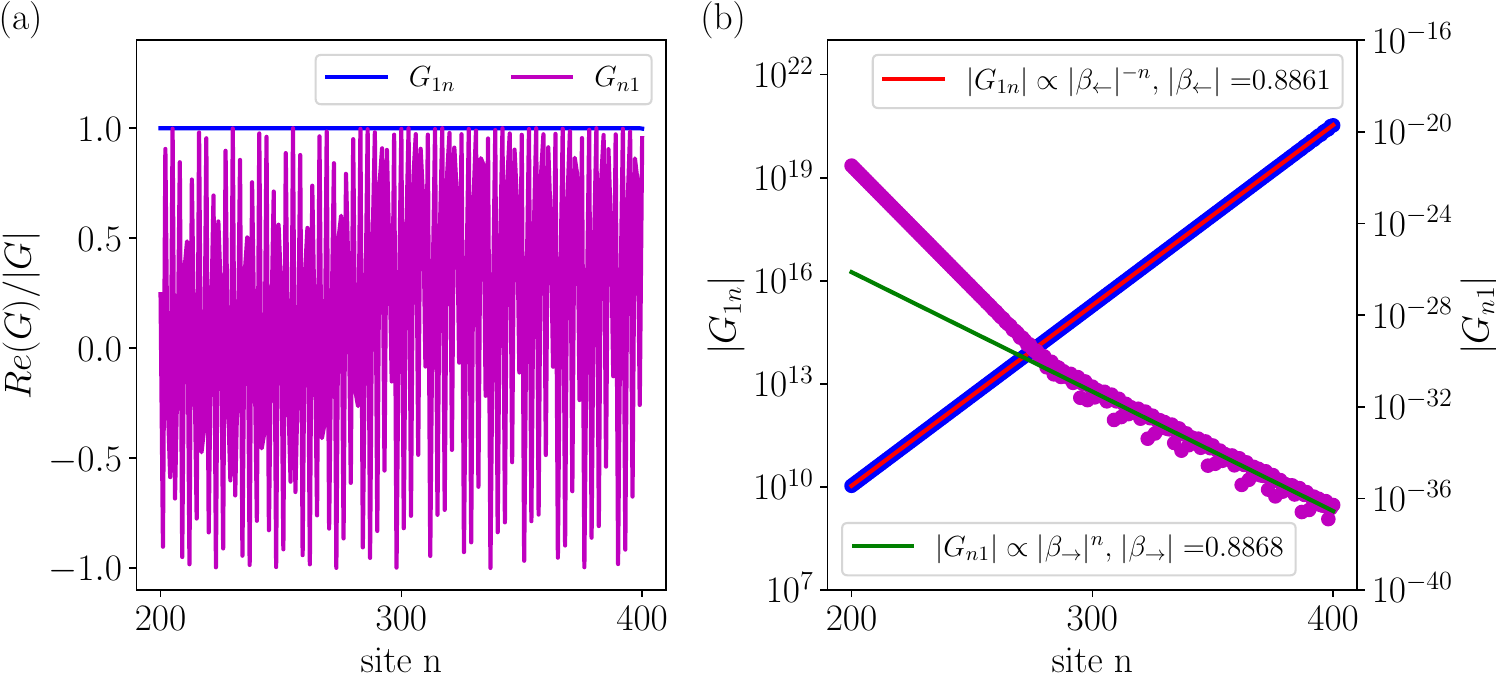}
    \caption{Behavior of the Green's function for a fixed system size. (a) Information on the phase of poles in Green's functions. We calculate $G_{n1}$ and $G_{1n}$, which encode the system’s responses in the forward and backward directions, respectively. (b) Scaling of the norm of the Green's function. The parameters are the same as those in Fig.~\ref{fig:GF_EM} (c) and (d). The scaling is consistent with the results shown in Fig.~\ref{fig:GF_EM} as $|\beta_{\leftarrow}|=0.8861\approx e^{-\alpha}, |\beta_{\rightarrow}| = 0.8868\approx e^{-\alpha}$.
    In the complex plane, $G_{1n}$ records the contribution from the poles outside the GBZ, while $G_{n1}$ records the contribution from the poles inside the GBZ. The absence of the oscillation in $G_{1n}$ indicates that there is a pole on the exterior of the GBZ with norm close to $e^{-\alpha}$, while the ``chaotic'' oscillation in $G_{n1}$ suggests there are multiple poles with norm close to $e^{-\alpha}$ but different phases in the interior of the GBZ. This provides firm evidence supporting our theoretical explanation.}
    \label{fig:gf_osc}
\end{figure}

In Fig.~\ref{fig:gf_osc}(a,b), we show the behavior of \(G_{1n}\) and \(G_{n1}\) for a fixed system size, with \(\omega\) chosen inside the circular region enclosed by the spectrum. In Fig.~\ref{fig:gf_osc}(a), we plot the phases of \(G_{1n}\) and \(G_{n1}\), represented by \(\operatorname{Re}G/|G|\). In Fig.~\ref{fig:gf_osc}(b), we plot \(|G_{1x}|\) and \(|G_{x1}|\). The scaling agrees with the results in Fig.~\ref{fig:GF_EM}(d), where we instead focus on \(G_{1L}/G_{L1}\) as the system size \(L\) is varied. Notably, the phase of \(G_{1n}\) remains consistently zero as \(n\) is varied, indicating that \(G_{1n}\) is controlled by a root of \(\omega-h(\beta)=0\) that lies very close to the real axis. This follows from the fact that \(G_{1n}\) is associated with the root inside the small loop of the squeezed GBZ near \(e^{-\alpha}\).

\section{\label{sec: 1D_Ronkin} Ronkin function formulation of squeezed GBZ in 1D}
In this section, we discuss how to understand the renormalization of the 1D GBZ using the Ronkin function. We first review some results about the Ronkin function for finite-range $h(\beta)$ in 1D non-Hermitian systems, which is defined as:

\begin{equation}\label{eq:ronkin_1D}
R_E(\mu) = \int_{0}^{2\pi}\frac{\mathrm{d}\theta}{2\pi}\,\log|f(e^{\mu+i\theta})|,
\end{equation}
where $f(\beta)=\det\left[E-h(\beta)\right]$. It takes the form $\det\left[E-h(\beta)\right] = a_{-M}(E)\beta^{-M}+\cdots+a_N(E)\beta^N$, the roots of which are $|\beta_1(E)|\le|\beta_2(E)|\le\cdots|\beta_{M+N}(E)|$. In this case, $R_E(\mu)$ takes a simple form \cite{Amoeba24}:

\begin{equation}\label{eq:ronkin_1D_2}
R_E(\mu) = \log|a_{-M}|-M\mu+\sum_{k = 1}^l(\mu-\log|\beta_k|),
\end{equation}
for $\log |\beta_l|\le\mu\le\log |\beta_{l+1}|$. Consequently, the Ronkin function is a piecewise linear function of $\mu$. The density of states can then be calculated through: 
\begin{equation}
\rho(E) = \frac{1}{2\pi}\nabla^2\phi(E),\quad \phi(E) = \min_{\mu\in\mathbb{R}} R_E(\mu).
\end{equation}

Meanwhile, the amoeba is defined as the set $\{\log|\beta|,\det\left[E-h(\beta)\right] = 0\}$. Consequently, the log of roots $\log|\beta_j|$s are components of the amoeba, and  intervals $[\log|\beta_j|,\log|\beta_{j+1}|]$ are amoeba holes. The central hole of the amoeba is the interval $[\log|\beta_M|,\log|\beta_{M+1}|]$, where the Ronkin function is flat: $\partial_{\mu} R_E(\mu) = -M+M = 0$. It has been shown \cite{Amoeba24} that an energy is in the spectrum ($\rho(E)\neq 0$) only if the central hole of the amoeba closes: $|\beta_M|=|\beta_{M+1}|$.

Now we turn to the non-Bloch Hamiltonian $h(\beta)$ with exponentially decaying long-range hopping:
\begin{align}
&h(\beta) = h_0(\beta)+\delta h(\beta),\nonumber\\
&\delta h(\beta) = g\sum_{n = 1}^L e^{-\alpha n}(\beta^n+\beta^{-n}).
\end{align}
We note that $h(\beta = e^{\mu+i\theta})$ diverges when $|\mu|>\alpha$, while $h(\beta)\approx h_0(\beta)$ for $-\alpha<\mu<\alpha$. As discussed in the main text, the density of states squeezed by long-range hoppings can be calculated with the domain of the Ronkin function restricted to $[-\alpha,\alpha]$:
\begin{equation}
\rho(E) = \frac{1}{2\pi}\nabla^2\phi(E),\quad \phi(E) = \min_{-\alpha\le\mu\le\alpha} R_E(\mu),
\label{eq:ronkin_1D_renormalized}
\end{equation}
where $R_E(\mu)$ is calculated using $h_0(\beta)$ when considering $g\to0$.

\begin{figure}[t]
    \centering
    \includegraphics[width=0.6\linewidth]{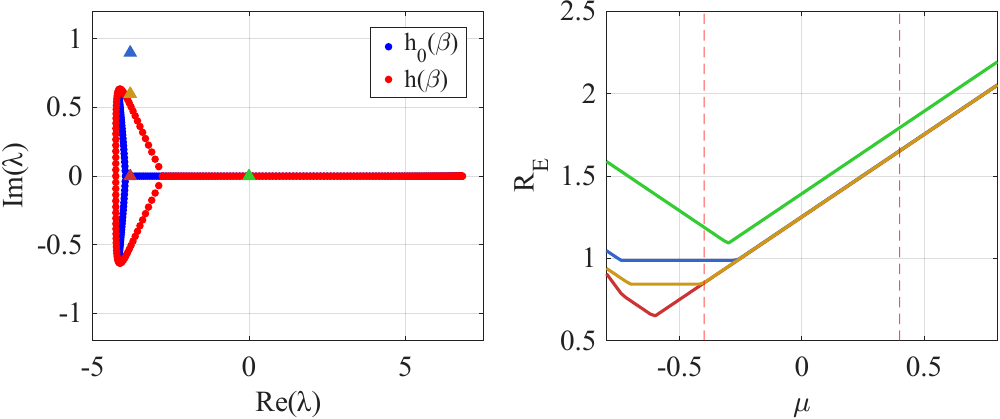}
    \caption{Ronkin function \(R_E(\mu)\) for the 1D system. (a) Energy spectra of the system with long-range hopping (\(h(\beta)\), red dots) and without long-range hopping (\(h_0(\beta)\), blue dots). (b) Ronkin functions \(R_E(\mu)\) for the four representative energies marked in panel (a). The red dashed lines indicate \(\mu=\pm\alpha\). If an energy lies in the spectrum of \(h_0\), the corresponding Ronkin function has a unique minimum at \(\mu_{\min}\); otherwise, a plateau appears. If the energy lies in the spectra of both \(h_0\) and \(h\), then \(|\mu_{\min}|<\alpha\) (green line). By contrast, if the energy lies in the spectrum of \(h\) but not in that of \(h_0\), a plateau appears whose endpoint is located at \(\mu=-\alpha\) (yellow line).}
    \label{fig:rk_1d}
\end{figure}

Different scenarios are illustrated in Fig.~\ref{fig:rk_1d}. In Fig.~\ref{fig:rk_1d}(a), we mark several representative energies by triangles of different colors. The corresponding Ronkin functions are shown in Fig.~\ref{fig:rk_1d}(b) using the same color scheme.

For the energy labeled by the green triangle, \(E\) lies in the spectra of both \(h_0\) and \(h\). The central hole of the amoeba closes, and the minimum of the Ronkin function is attained within the region \([-\alpha,\alpha]\). This corresponds to the fact that the portion of the GBZ of \(h_0\) lying inside the annulus \(e^{-\alpha}<|\beta|<e^{\alpha}\) remains part of the squeezed GBZ of \(h\).

For the energy labeled by the red triangle, \(E\) lies in the spectrum of \(h_0\) but not in the spectrum of \(h\). The central hole of the amoeba closes, but the minimum of the Ronkin function is attained outside the region \([-\alpha,\alpha]\). Within \([-\alpha,\alpha]\), the minimum is instead reached at \(\mu=-\alpha\). Using Eq.~\eqref{eq:ronkin_1D} and Eq.~\eqref{eq:ronkin_1D_renormalized}, one finds
\[
\rho(E)=\frac{1}{2\pi}\nabla^2 R_E(-\alpha)=0.
\]
This corresponds to the fact that the portion of the GBZ of \(h_0\) outside the annulus \(e^{-\alpha}<|\beta|<e^{\alpha}\) does not remain on the squeezed GBZ of \(h\).

For the energy labeled by the yellow triangle, \(E\) lies in the spectrum of \(h\) but not in the spectrum of \(h_0\). The central hole of the amoeba on \(\mathbb{R}\) does not close; however, it terminates exactly at \(\mu=-\alpha\). The minimum of the Ronkin function within the region \([-\alpha,\alpha]\) is attained at \(\mu=-\alpha\), which now lies on the squeezed amoeba. Using Eq.~\eqref{eq:ronkin_1D}\eqref{eq:ronkin_1D_renormalized}, one obtains
\[
\rho(E)=\frac{1}{2\pi}\nabla^2 R_E(-\alpha)\neq 0,
\]
showing that \(E\) belongs to the spectrum of \(h\). This corresponds to the fact that the portion of the GBZ of \(h_0\) outside the annulus \(e^{-\alpha}<|\beta|<e^{\alpha}\) is squeezed to the corresponding segment on \(|\beta|=e^{\pm\alpha}\).

Finally, for the energy labeled by the blue triangle, \(E\) lies in neither the spectrum of \(h_0\) nor that of \(h\). The central hole of the amoeba is open and has a finite overlap with \([-\alpha,\alpha]\). Thus, \(E\) is clearly not in the spectrum of \(h_0\). For \(h\), the minimum of the Ronkin function on \([-\alpha,\alpha]\) is the same as that on $\mathbb{{R}}$. In this case,
\[
\rho(E)=\frac{1}{2\pi}\nabla^2 R_E(-\alpha)=0,
\]
and \(E\) is not in the spectrum of \(h\).

\section{\label{sec: chiral_sm} Squeezed GBZ in Lindbladian open quantum systems}
\begin{figure}[t]
    \centering
    \includegraphics[width=0.9\linewidth]{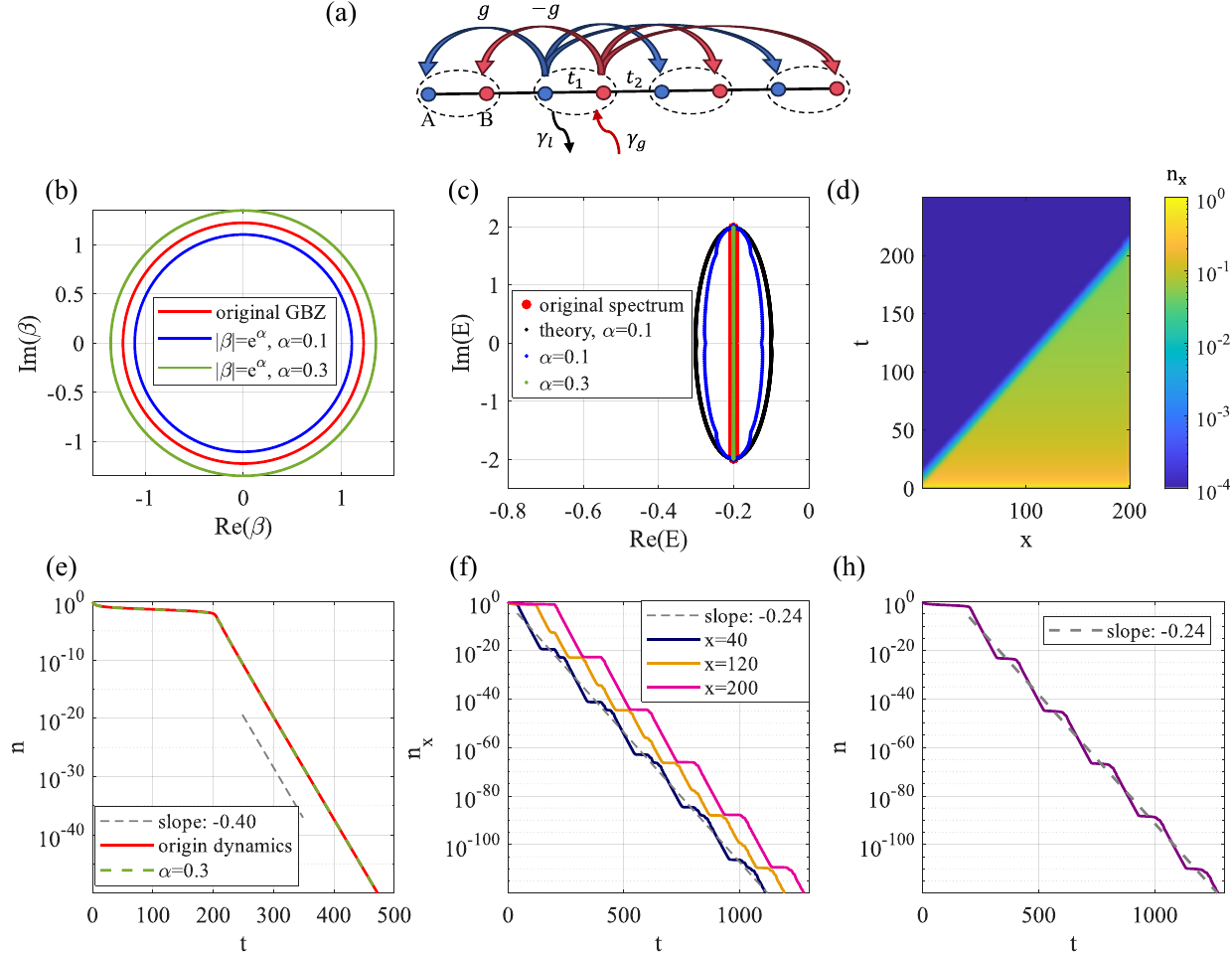}
    \caption{Long-range hopping reshapes the chiral damping dynamics.
(a) Schematic of the model with long-range hopping, $\mathcal{L} = \mathcal{L}_0+\mathcal{L}_1$ in Eq.~\eqref{eq:chiral_L0} and Eq.~\eqref{eq:H1}. (b) Generalized Brillouin zone (GBZ) of the original model $\mathcal{L}_0$ (red). The green and blue circles $|\beta| = \exp(\alpha)$ for $\alpha = 0.1$ and $0.3$, respectively, where $\alpha$ is the exponential decay rate of the long-range hopping. (c) Open-boundary spectrum. Red: spectrum of $\mathcal{L}_0$. Black: spectrum derived from the squeezed GBZ. Blue: spectrum with long-range hopping at $\alpha = 0.1$, which closely follows the black spectrum. Green: spectrum with $\alpha = 0.3$, preserving the original spectral structure. (d) Early dynamics of the local density $n_x = \sum_{s=A,B}c_{xs}^{\dagger}c_{xs}-n_x^s$ for $\alpha = 0.1$, where $n_x^s$ is the density of steady state. (e) Dynamics of the site-averaged fermion number \(n = \sqrt{\sum_x n_x^2 / L}\). For \(\alpha=0.3\), the dynamics are essentially identical to the original case: on timescales proportional to the system size they are gapless (reflecting the gapless PBC spectrum), while on longer timescales they decay exponentially with a coefficient \(0.40\), consistent with the OBC Liouvillian gap \(\Lambda = \min_n[2\Re(-\lambda_n)]\) in (c).  
(f), (h) Local density dynamics and site-averaged fermion number for \(\alpha=0.1\). The evolution exhibits a series of periodic plateaus, and at long times an exponential decay whose rate matches the new OBC Liouvillian gap seen in (c).  
Common parameters: \(t_1 = 1.0\), \(t_2 = 1.0\), \(\gamma_l = 0.2\), \(\gamma_g = 0.2\), \(g = 0.01\), \(L = 200\).}
    \label{fig:chiral_sm}
\end{figure}

To investigate how long-range hopping reshapes dissipative quantum systems, we study a one-dimensional chain with two sublattices A and B, following Ref.~\cite{song2019non} and as illustrated in Fig.~\ref{fig:chiral_sm}(a). The dynamics is governed by the Lindbladian
\begin{eqnarray}
\label{eq:chiral_L0}
    &&H = \sum_{ij}h_{ij}c_i^{\dagger}c_j+\mathrm{h.c.} = \sum_{x}\bigl(t_1c_{xA}^{\dagger}c_{xB}+t_2c_{xB}^{\dagger}c_{x+1A}\bigr)+\mathrm{h.c.},\nonumber\\
    &&L_{\mu}^l = \sum_iD_{\mu i}^lc_i = \sqrt{\gamma^l/2}\,(c_{xA}-ic_{xB}),\quad L_{\mu}^g = \sum_{i}D_{\mu i}^g c_i^{\dagger} = \sqrt{\gamma^g/2}\,(c_{xA}^{\dagger}+ic_{xB}^{\dagger}),
    \nonumber\\
    &&\mathcal{L}_0[\rho] = -i[H, \rho]+\sum_{\mu}\bigl(2L_{\mu}^l\rho L_{\mu}^{l\dagger} - \{L_{\mu}^{l\dagger}L_{\mu}^l, \rho\}\bigr)+\sum_{\mu}\bigl(2L_{\mu}^g\rho L_{\mu}^{g\dagger} - \{L_{\mu}^{g\dagger}L_{\mu}^g, \rho\}\bigr),
\end{eqnarray}
where $i = xs,\; s=A,B$. On top of this, we introduce an additional exponentially decaying long-range hopping term in the Hamiltonian,
\begin{equation}
\label{eq:H1}
    H_1 = \sum_{x\ne x'}g\exp(-\alpha|x-x'|)(c_{xA}^{\dagger}c_{x'A}-c_{xB}^{\dagger}c_{x'B})+\mathrm{h.c.},\quad \mathcal{L}_1[\rho]=-i[H_1, \rho]. 
\end{equation}
We study the evolution of the single-particle correlation function $\Delta_{ij} = \tr[c_i^{\dagger}c_j\rho]$ under the total Lindbladian $\mathcal{L}_0+\mathcal{L}_1$, whose dynamics is described by
\begin{equation}
\label{Eq:Chiral}
    \frac{d\Delta}{dt} = X\Delta+\Delta X^{\dagger}+2M_g,\quad X = i h^{\mathrm{T}}-(M_l^{\mathrm{T}}+M_g),\quad (M_g)_{ij} = \sum_{\mu}D_{\mu i}^{g*}D_{\mu j}^g,\quad (M_l)_{ij} = \sum_{\mu}D_{\mu i}^{l*}D_{\mu j}^l.
\end{equation}
The exact steady state is known as $\Delta_s = (\gamma_g/(\gamma_l+\gamma_g))I_{2L\times 2L}$. 
However, the dynamics towards the steady state is intriguing and governed by the non-Hermitian matrix $X$ through the deviation $\tilde{\Delta} = \Delta-\Delta_s$, which obeys
\begin{equation}
    \frac{d\tilde{\Delta}}{dt} = X\tilde{\Delta}+\tilde{\Delta}X^{\dagger}.
\end{equation}
Note that the long-range hopping in Eq.~\eqref{eq:H1} has opposite signs for the A and B sublattices. This choice preserves chiral symmetry, so that the spectrum of $X$ remains symmetric with respect to the real axis, allowing us to cleanly isolate the effects of long-range hopping. Our main results below, however, are independent of the specific choice of the hopping strengths.

Without long-range hopping and under periodic boundary conditions (PBC), $X(k)$ takes the form
\begin{equation}
    X(k) = i\bigl[(t_1+t_2\cos k)\sigma_x+(t_2 \sin k-i\tfrac{\gamma}{2})\sigma_y\bigr]-\frac{\gamma}{2}I,\quad \gamma = \gamma_l+\gamma_g.
\end{equation}
The Liouvillian gap is defined as $\Lambda = \min_n[2\Re(-\lambda_n)]$, where $\lambda_n$ are the eigenvalues of $X$. For $t_1\le t_2$, the gap closes ($\Lambda = 0$), while for $t_1>t_2$ a finite gap opens. Under open boundary conditions (OBC), the spectrum of $X$ is described by the non-Bloch theory with the generalized Brillouin zone (GBZ) introduced in the main text. Here the GBZ is a circle of radius $|\beta| = \exp(-\kappa)$, where $\kappa = -\ln\sqrt{|(t_1+\gamma/2)/(t_1-\gamma/2)|}$, as shown in Fig.~\ref{fig:chiral_sm}(b). In what follows, we focus on the parameters $t_1 = t_2 = 1.0$ and $\gamma_l = \gamma_g = 0.2$, giving $\kappa \approx -0.2$. Remarkably, although the system remains gapless under PBC, a finite OBC gap opens, as seen in Fig.~\ref{fig:chiral_sm}(c).

When long-range hopping is added, the eigenstates under PBC remain extended Bloch waves, confirming that $H_1$ acts as a perturbation as long as $g\ll t_1,t_2,\gamma$. Under OBC, however, when the GBZ of $\mathcal{L}_0$ lies outside the annulus bounded by $e^{-\alpha}$ and $e^{\alpha}$, the system's GBZ is replaced by the squeezed GBZ discussed in the main text. Here the squeezed GBZ is simply the circle $|\beta| = \exp(\alpha)$ for $\alpha < -\kappa$, while the GBZ is essentially unchanged for $\alpha \ge -\kappa$. Consequently, in Fig.~\ref{fig:chiral_sm}(c) the OBC spectrum for $\alpha = 0.3$ (green) coincides with the original spectrum (red), whereas for $\alpha = 0.1$ (blue) it forms a loop very close to the theoretical squeezed GBZ (black). 

The changes of the spectrum directly impact the dynamics of the systems. 
For simulation, we chose the initial state as $\tilde{\Delta}(t=0) = (\gamma_l/\gamma) I_{2L\times 2L}$.
Figure~\ref{fig:chiral_sm}(d) shows the early dynamics of the local density $n_x = \sum_{s=A,B}c_{xs}^{\dagger}c_{xs}-n_x^s$, where $n_x^s$ is the steady-state density. The dynamics clearly exhibits chiral damping: a sharp wavefront separates a region of polynomial decay (dictated by the gapless PBC spectrum) from a region of exponential decay (controlled by the gapped OBC spectrum). At late times, if $\alpha>-\kappa$, the long-range hopping does not alter the original dynamics because both the GBZ and the spectrum remain unchanged [Fig.~\ref{fig:chiral_sm}(e)]. In contrast, when $\alpha<-\kappa$ [Fig.~\ref{fig:chiral_sm}(f)], a series of plateaus emerge in the dynamics, and on long time scales the system displays a slower exponential decay whose rate is determined by the new OBC Liouvillian gap shown in Fig.~\ref{fig:chiral_sm}(c). The plateaus in dynamics have a twofold understandings: they are linked to the oscillations associated with conjugate pairs of eigenvalues in the OBC spectrum, and they can be understood in a simple physical picture as the wavefront propagating to the boundary and subsequently returning to the bulk on a time scale set by the long-range hopping, thereby sustaining the chiral damping dynamics.

We should also note that for a quadratic fermionic Lindblad equation whose single-particle
correlation matrix obeys Eq.~\eqref{Eq:Chiral}, the spectrum of the Liouvillian is directly related to the spectrum of $X$,
where $X$ is an $N\times N$ matrix. We denote its eigenvalues, including
algebraic multiplicities, by
\begin{equation}
  \spec(X)=\{x_i\}_{i=1}^{N}.
  \label{eq:X-spectrum}
\end{equation}

For the complete quadratic problem, the Liouvillian can be brought to a
normal-master-mode form away from defective points. In the convention used
here, its $2N$ elementary rapidities are
\begin{equation}
  \mathcal R
  =\{x_1,\ldots,x_N,x_1^*,\ldots,x_N^*\}.
  \label{eq:rapidities}
\end{equation}
Equivalently, the diagonal normal form may be written spectrally as
\begin{equation}
  \mathcal L
  \sim
  \sum_{i=1}^{N}\left(x_i\hat n_i+x_i^*\hat m_i\right),
  \label{eq:normal-form}
\end{equation}
where the mutually commuting normal-mode occupation operators have binary
eigenvalues $n_i,m_i\in\{0,1\}$. It follows directly that every Liouvillian
eigenvalue is a subset sum of the elementary rapidities,
\begin{equation}
  \lambda_{\boldsymbol n,\boldsymbol m}
  =\sum_{i=1}^{N}\left(n_i x_i+m_i x_i^*\right),
  \qquad n_i,m_i\in\{0,1\}.
  \label{eq:full-spectrum}
\end{equation}
There are $2^{2N}=4^N$ binary occupation patterns and hence $4^N$
Liouvillian eigenvalues counting algebraic multiplicity. The empty pattern, $n_i=m_i=0$ for all $i$, gives the stationary
eigenvalue $\lambda=0$.

These results indicate that the influence of long-range hopping in non-Hermitian systems, can be directly generalized to the dissipative quantum system for the nontrivial dynamics. Both the OBC spectrum and the associated dynamical behavior in the presence of long-range hopping can be fully captured by the squeezed GBZ, which therefore provides a unified description for this class of open quantum systems.

\bibliography{references}